\documentclass[authoryear,final,3p,times]{elsarticle}
\usepackage{amssymb}
\usepackage{amsthm}
\usepackage{amsmath}
\usepackage{bm}
\usepackage{array}
\usepackage{ulem}
\usepackage{longtable}
\usepackage{CJK}
\usepackage{setspace}
\usepackage{multirow}
\usepackage{siunitx}
\usepackage{threeparttable}
\usepackage{verbatim}
\usepackage{natbib}
\usepackage[justification=centering]{caption}
\usepackage{lineno}
\usepackage{longtable}
\usepackage[breaklinks,colorlinks,linkcolor=black,citecolor=black,urlcolor=black]{hyperref}
\usepackage{subfigure}
\usepackage{color}
\begin{document}
\begin{sloppypar}
\hypersetup{CJKbookmarks=true}
\begin{CJK*}{GBK}{song}
\begin{frontmatter}
    \title{Center-to-face momentum interpolation and face-to-center flux reconstruction in Euler-Euler simulation of gas-solid flows}
    \author[label1,label2]{Yige Liu}
    \author[label1,label3]{Bidan Zhao}
    \author[label1]{Ji Xu}
    \author[label1,label3]{Junwu Wang\corref{cor1}}
    \cortext[cor1]{Corresponding author}
    \ead{jwwang@cup.edu.cn}
    \address[label1]{State Key Laboratory of Mesoscience and Engineering, Institute of Process Engineering, Chinese Academy of Sciences, Beijing 100190, P. R. China}
    \address[label2]{School of Chemical Engineering, University of Chinese Academy of Sciences, Beijing, 100049, P. R. China}
    \address[label3]{College of Mechanical and Transportation Engineering, China University of Petroleum, Beijing, 102249, P. R. China}
    \begin{abstract}
    In order to resolve the pressure checkerboard field problem with collocated grid, it is essential to employ the momentum interpolation method when formulating the pressure equation, and the flux reconstruction method when updating the cell-centered velocity fields.
    In this study, we first derive a momentum interpolation method for Euler-Euler simulation of gas-solid flows, which is independent of the time step, the transient term discretization scheme, the under-relaxation factor and the shape of grid; a complete first-order flux reconstruction method is then proposed to update the cell-centered velocities. Their effectiveness are proved by simulating the hydrodynamics of solids settlement, gas-solid fixed bed, bubbling fluidized bed and circulating fluidized bed riser, and then comparing the simulation results to the theoretically known solutions. Their superiority over the standard solver of OpenFOAM$^\circledR$ in suppressing the high-frequency oscillations and enhancing the smoothness and accuracy is also proved.
    Finally, the difficulty in fully eliminating the high-frequency oscillations is attributed to the insufficiency of current methods in
    handling the situations where the independent variables undergo abrupt change.
    \end{abstract}
\end{frontmatter}

\section{Introduction}
Two-phase flows, characterized by the presence of two phases such as gas and solids, are encountered in a wide range of industrial and natural processes. The complex interactions between two phases make the analysis and prediction of two-phase flow behavior challenging.
With the development of computer technology, computational fluid dynamics (CFD) simulation has become an essential component in the research and development of two-phase flow systems. Its ability to provide detailed insights into the complex behavior of two-phase flows, coupled with its cost-effectiveness and versatility, makes it a promising technique for addressing the challenges associated with these flows. As the field of CFD continues to evolve, its applications in the study of two-phase flows are expected to expand, offering new opportunities for improving industrial processes and solving real-world problems \citep{prosperetti2009computational,yeoh2019computational,wang2020continuum}.

The collocated variable arrangements, where the vector components and scalar variables are all saved at the same physical location, are the mainstream choice for state-of-the-art CFD \citep{darwish2016finite,ferziger2019computational}. Due to its versatility and efficiency, it is generally preferred over the staggered variable arrangements \citep{harlow1965numerical}, where vector components and scalar variables are stored at separate physical locations. Collocated strategy is well-suited for engineering applications with non-Cartesian grid and unstructured grid \citep{demirdzic1982finite,hsu1982curvilinear,peric1985finite}, and require less memory overhead with coding clarity. In addition, multigrid and multilevel techniques, which significantly decrease computational costs on fine grids, are also more readily applicable to a collocated arrangement \citep{darwish2016finite}.
Nevertheless, there is a checkerboard issue \citep{patankar2018numerical} that arises with the gas pressure field when central differencing scheme is employed in both the continuity equations and the gas pressure gradient in the momentum equations. This plight allows for the coexistence of two distinct gas pressure fields, leading to the nonphysical oscillations of gas pressure. As a consequence, the gradient of gas pressure within the current grid is contingent upon the gas pressure levels at the two alternating, non-consecutive grid points that straddle the grid. The same is applicable to the continuity equations, which only preserve conservation for alternating grids.

To get over with the checkerboard pressure field or the decoupling of the velocity and pressure in collocated variable arrangements, the well-known original momentum interpolation method (OMIM) was proposed \citep{rhie1983numerical} and validated by \citet{peric1988comparison} for steady-state flows with no relaxation. For structured grids, physically consistent arguments combined with the Rhie-Chow interpolation method yields results that are consistent with staggered variable arrangements at lower computational burden and smaller storage space \citep{peric1988comparison}.
However, the converged solution of the OMIM is dependent on the under relaxation factor used and hence particular handling has been taken to remove this dependency \citep{majumdar1988role,miller1988use}. This kind of correction term can be applied to any pressure-velocity coupled algorithm without considerably affecting the convergence rate \citep{martinez2017influence}. Nevertheless, to yield accurate results, it needs to be paired with a fine enough mesh.
For unsteady flows, OMIM is also dependent on the time step used \citep{ferziger2019computational,kawaguchi2002checkerboard}. Including a transient correction term could resolve this dependency \citep{choi1999note,bartholomew2018unified,xiao2017fully}. Various time discretization schemes have been investigated \citep{shen2001improved,cubero2007compact}. Additionally, new techniques in momentum interpolation that do not rely on both the time step and the under-relaxation factor have also been proposed \citep{yu2002discussion,cubero2007compact,pascau2011cell,udaykumar1997multiphase,issa1994numerical}. Moreover, a consolidated formation for the momentum interpolation method has been developed on random grids, which includes extensions for large discontinuous source terms and discontinuous changes in density \citep{bartholomew2018unified}. Furthermore, Rhie-Chow momentum interpolation methods have been coupled into algorithms where the pressure corrector step is discarded and the velocity and pressure are solved at the same time \citep{chen2010coupled,denner2014fully} and sharp-interface immersed boundary method \citep{yi2016improved,tseng2003ghost,mittal2008versatile,udaykumar1997multiphase}. The previous advancements of the momentum interpolation method primarily concentrated on the field of single-phase flow. Although \cite{bartholomew2018unified} extended the applicability of the aforementioned method to encompass two-phase flow with disparate densities, the interaction forces between the phases such as drag force and the two-phase flow pertaining to the solid phase were not taken into account.
\citet{cubero2014consistent} extended the momentum interpolation method from single-phase flows to multiphase flows, which considers the existence of transient terms and includes a correction for the drag interaction amid phases. However, the momentum interpolation methods that are suitable for the situation of multiphase flows related to the solid phase are not discussed. Consequently, there is a dearth of a unified momentum interpolation methods that are applicable to two-phase flows, with due consideration for the solid phase where the solid pressure should be taken into account.

Rhie-Chow momentum interpolation method can solve the problem of velocity-pressure decoupling in collocated grids by constructing conservation equations at the interfaces between cells, which is concerned with the interpolation process from the cell center to the face center. While the flux reconstruction process from the face center to the cell center, which is the inverse of momentum interpolation process, deserves to be paid attention to as well. Because inappropriate methods can cause high-frequency oscillations of the velocities.
The form of reconstruction operator can be determined by solving a flux error minimization problem \citep{shashkov1998local,weller2014non,weller2014curl} or calculating a volume weighted average \citep{10.1007/0-387-38034-5_9}.
In addition, in the two-phase flow simulation using the collocated grid, the velocities are generally updated by partial reconstruction. That is, the reconstruction part only includes the shared pressure term and part of the source term \citep{weller2002derivation,rusche2003computational}, which is widely used by \citet{passalacqua2011ImplementationIterativeSolution,liu2014cfd,venier2016numerical}.
In this manner, the high-frequency oscillation of the unreconstructed part of the flux cannot be filtered during the velocity update. The reconstructed object is not the complete flux obtained by the momentum interpolation method, rather, it is merely a portion of the flux. Therefore, the smoothness property of the momentum interpolation method cannot be well maintained during the flux reconstruction step.
\citet{aguerre2018oscillation} proposed a new approach that based on a flux error minimization problem to address the issue of spurious high-frequency in cell-centered velocities. This approach takes full advantage of the non-oscillation quality of the interfacial flux and it has been verified to guarantee second-order accuracy in single-phase flow. To date, this method has not been extended to two-phase flows and has not been combined with a suitable momentum interpolation method.

In this article, improved momentum interpolation method and flux reconstruction method are proposed in order to address the center-to-face momentum interpolation and face-to-center flux reconstruction problems met in the state-of-the-art Euler-Euler simulation of gas-solid flows with collocated grids. Extensive simulations are then carried out to prove their effectiveness and their superiority over the standard OpenFOAM$^\circledR$ solver. 

\section{Discretized momentum equations}
\begin{longtable}[H]{lr}
    \caption{Governing equations. \label{tab:Gov}}
    \\\hline
    Mass conservation equations \\
    $\frac{\partial}{\partial t}\left(\alpha_g\rho_g\right)+\nabla\cdot\left(\alpha_g\rho_g\bm u_g\right)=0$ \\
    $\frac{\partial}{\partial t}\left(\alpha_s\rho_s\right)+\nabla\cdot\left(\alpha_s\rho_s\bm u_s\right)=0$ \\
    Momentum conservation equations \\
    $\frac{\partial }{\partial t}\left(\alpha_g\rho_g\bm u_g\right)+\nabla\cdot\left(\alpha_g\rho_g\bm u_g\bm u_g\right)=-\alpha_g\nabla p_g+\nabla\cdot\left(\alpha_g\bm \tau_g\right)+\alpha_g\rho_g\bm g-\bm F_{drag}$ \\
    $\frac{\partial }{\partial t}\left(\alpha_s\rho_s\bm u_s\right)+\nabla\cdot\left(\alpha_s\rho_s\bm u_s\bm u_s\right)=-\alpha_s\nabla p_g -\nabla p_s+\nabla\cdot\left(\alpha_s\bm \tau_s\right)+\alpha_s\rho_s\bm g+\bm F_{drag}$ \\
    \hline
\end{longtable}
Table. \ref{tab:Gov} presents the governing equations of the two-fluid model that are used in this work, where the subscript $g$ denotes the gas phase and the subscript $s$ denotes the solid phase. $\alpha_g$ and $\alpha_s$ are the gas-phase and solid-phase volume fractions. $\rho_g$ and $\rho_s$ are the gas-phase and solid-phase densities. $\bm u_g$ and $\bm u_s$ are the gas-phase and solid-phase velocities. $p_g$ is the gas pressure and $p_s$ is the granular pressure. $\tau_g$ and $\tau_s$ are the gas-phase and solid-phase stress tensors. $\bm g$ is the gravitational acceleration and $F_{drag}=\beta(\bm u_g - \bm u_s)$ is the interphase drag force. Details of the constitutive relations used are summarized in Appendix A.
According to Table \ref{tab:Gov}, the $k$-phase momentum equation is
\begin{equation}
    \frac{\partial }{\partial t}\left(\alpha_k\rho_k\bm u_k\right)+\beta \bm u_k+\nabla\cdot\left(\alpha_k\rho_k\bm u_k\bm u_k\right)=\nabla\cdot\left(\alpha_k\bm \tau_k\right)-\alpha_k\nabla p_g - \zeta \nabla p_s +\alpha_k\rho_k \bm g+\beta \bm u_{1-k},
    \label{equ:RCtfms}
\end{equation}
where $\zeta=1$ for solid phase and $\zeta=0$ for gas phase, the subscript $k$ denotes the current phase and $1-k$ denotes another phase.
If solving the transient term with a first-order backward Euler scheme, the $k$-phase momentum equation is
\begin{equation}
\begin{split}
\frac{\alpha_k \rho_k \bm u_k}{\Delta t} + \beta \bm u_k + \nabla\cdot\left(\alpha_k\rho_k\bm u_k\bm u_k\right)=\nabla\cdot\left(\alpha_k\bm \tau_k\right) + \frac{\alpha_k \rho_k \bm u_k^i}{\Delta t}
-\alpha_k\nabla p_g - \zeta \nabla p_s +\alpha_k\rho_k \bm g+ \beta \bm u^*_{1-k},
\end{split}
\end{equation}
where superscript $i$ denotes the previous time step, superscript $*$ denotes the previous iteration.
Using the finite volume method, the semi-discretized formulation is
\begin{equation}
\mathbb A_k \bm u_k = \bm b_k + \frac{\alpha_k \rho_k \bm u_k^i}{\Delta t} -\alpha_k \nabla p_g - \zeta\nabla p_s+\alpha_k \rho_k \bm g + \beta\bm u^*_{1-k},
\end{equation}
where $\mathbb A_k$ is the coefficients matrix of the momentum equation including the contribution of transient term, convection term, diffusion term and the implicit part of drag term. $\bm b_k$ is the corresponding source term from which the transient source term is extracted. Then we have
\begin{equation}
\mathbb D_k \bm u_k = \mathbb H_k + \frac{\alpha_k \rho_k \bm u_k^i}{\Delta t}-\alpha_k \nabla p_g - \zeta\nabla p_s+\alpha_k \rho_k \bm g + \beta\bm u^*_{1-k},
\end{equation}
where $\mathbb D_k$ is the diagonal part of the momentum coefficients matrix $\mathbb A_k$ and $\mathbb H_k=\bm b_k - ({\mathbb A_k-\mathbb D_k}) \bm u_k^{*}$ is introduced by Jacob iteration scheme, which provides a simple way of obtaining an approximate solution of phase velocity \citep{passalacqua2011ImplementationIterativeSolution}.

If the under relaxation technique is applied for $\mathbb D_k \bm u_k = \mathbb H_k + \frac{\alpha_k \rho_k \bm u_k^i}{\Delta t}$, the remaining source terms $- \zeta\nabla p_s+\alpha_k \rho_k \bm g + \beta\bm u^*_{1-k}$ should be added after the relaxation process completed. Then the diagonal momentum coefficients will be divided into two terms: the part combined with the current velocity to be solved $\frac{\mathbb D_k}{\lambda}$ and the part combined with the previous iteration velocity $(1-\frac{1}{\lambda}) {\mathbb D_k}$, where $\lambda$ is the under relaxation factor lies between 0 and 1. A smaller value of $\lambda$ will result in a smaller fluctuation of the iteration value, but the convergence rate will be reduced. Thus we have $\frac{1}{\lambda}\mathbb D_k \bm u_k = \mathbb H_k + \frac{\alpha_k \rho_k \bm u_k^i}{\Delta t}+ \frac{1-\lambda}{\lambda} \mathbb D_k \bm u_k^*$, which is equivalent to $\mathbb D_k \bm u_k = \lambda \mathbb H_k + \lambda \frac{\alpha_k \rho_k \bm u_k^i}{\Delta t}+(1-\lambda) \mathbb D_k \bm u_k^*$.
Then the under-relaxed $k$-phase momentum equation becomes
\begin{equation}
\begin{split}
\bm u_k = \lambda \frac{\mathbb H_k}{\mathbb D_k} + \frac{\lambda}{\Delta t} \frac{\alpha_k \rho_k \bm u_k^i}{\mathbb D_k}+ (1-\lambda)\bm u_k^*-\frac{\alpha_k}{\mathbb D_k} \nabla p_g - \zeta\frac{1}{\mathbb D_k}\nabla p_s+\frac{\alpha_k \rho_k}{\mathbb D_k} \bm g + \frac{\beta}{\mathbb D_k} \bm u^*_{1-k}.
\label{equ:Momentum}
\end{split}
\end{equation}

\section{Generalized momentum interpolation method}
\begin{figure}[!htb]
    \centering
    \includegraphics[scale=1]{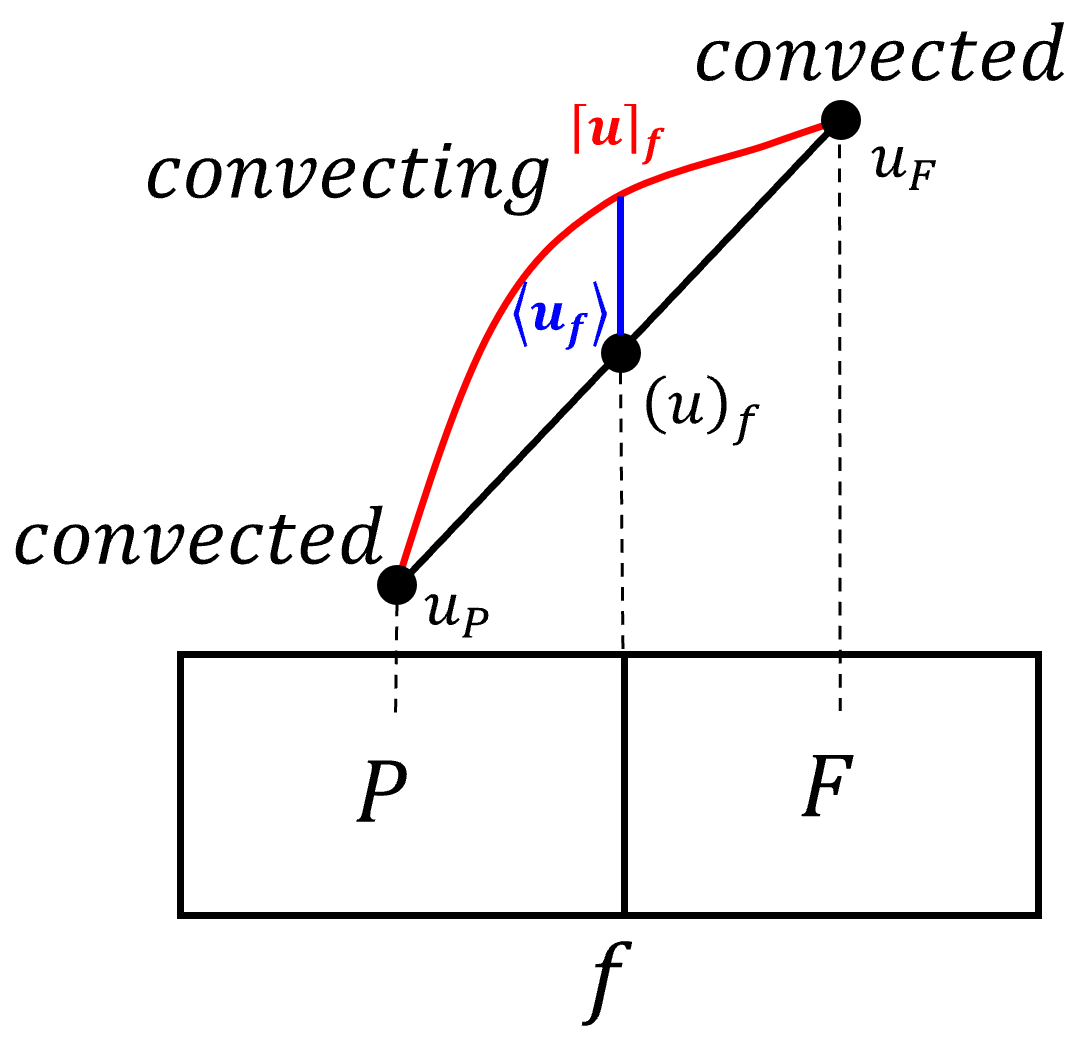}
    \caption{Schematic of momentum interpolation method.}
    \label{grid}
\end{figure}

In order to solve the problem of pressure-velocity decoupling, the Rhie-Chow interpolation method mimics the staggered grid formulation by forming a pseudo-momentum equation to obtain the convecting values at the face center $f$ instead of merely by linear interpolation. It is because of this behavioral imitation that the Rhie-Chow interpolation is successful.
Thus, according to momentum interpolation method, the generic form for values at the face can be expressed as
\begin{equation}
\begin{split}
\lceil \odot \rceil _f = (\odot)_f + \langle \odot \rangle _f,
\label{equ::RC}
\end{split}
\end{equation}
where the delimiter $\lceil \odot \rceil _f$ denotes the convecting value of $\odot$ at the face center $f$ calculated by pseudo-momentum equation, the delimiter $(\odot)_f$ denotes the linear interpolation, the delimiter $\langle \odot \rangle _f$ denotes the correction to the linear interpolation.
The linear interpolation usually uses the inverse distance weighting method \citep{greenshieldsweller2022} through neighbouring convected values $(\odot)_P$ and $(\odot)_F$ in the cell center that shares the same face $f$
\begin{equation}
(\odot)_f = \omega (\odot)_P + (1-\omega) (\odot)_F,
\label{linear}
\end{equation}
where $\omega = \frac{\overline{fF}}{\overline{fF}+\overline{fP}}$ is the inverse distance weighting factor related to the position of the face $f$ with respect to the cell center $P$ and $F$. $\overline{fF}$ and $\overline{fP}$ are the distances between the cell center $F$ and the face center $f$ and between the cell center $P$ and the face center $f$, respectively.

In accordance with the fundamental tenet of momentum interpolation method, Eq. (\ref{equ::RC}) can be extended to the gas-solid two-phase flow system employed in this study, resulting in the following final form (see Appendix B for detailed derivation):
\begin{equation}
\begin{split}
\lceil \bm u_{k} \rceil _f = (\bm u_k)_f+ \frac{\lambda}{\Delta t}\left(\frac{\alpha_k \rho_k}{\mathbb D_k}\right)_f\left[\lceil \bm u_{k}^i \rceil _f - (\bm u^i_k)_f\right] + (1 -\lambda)\left[\lceil \bm u_{k}^* \rceil _f - (\bm u^*_k)_f\right]\\
-\left( \frac{\alpha_{k}}{\mathbb D_k}\right)_f \left[\lceil \nabla p_g \rceil _f- \left( \nabla p_g \right)_f \right]- \zeta \left(\frac{1}{\mathbb D_k}\right)_f\left[ \lceil \nabla p_s \rceil_f - \left( \nabla p_s \right)_f \right]
+\left(\frac{\beta}{\mathbb D_k}\right)_f\left[ \lceil \bm u_{1-k}^* \rceil _f - (\bm u_{1-k}^*)_f\right].
\label{RCukf2}
\end{split}
\end{equation}

Eq. (\ref{RCukf2}) can also be expressed as

\begin{equation}
\begin{split}
&\lceil \bm u_{s} \rceil _f = \\
&\left(\bm u^{pre}_{s}\right)_f + \frac{\lambda}{\Delta t}\left(\frac{\alpha_s \rho_s}{\mathbb D_s}\right)_f \left[\lceil \bm u^i_{s} \rceil _f - \left( \bm u^i_s \right)_f \right]
+ (1 -\lambda)\left[\lceil \bm u^*_{s} \rceil _f
    - \left(\bm u^*_s \right)_f \right]
     - \left(\frac{1}{\mathbb D_s}\right)_f\nabla^{\perp}p_s
    - \left(\frac{\alpha_s}{\mathbb D_s}\right)_f \nabla^{\perp}p_g + \left(\frac{\alpha_s\rho_s}{\mathbb D_s}\right)_f \bm g
    +\left(\frac{\beta}{\mathbb D_s}\right)_f \lceil \bm u_{g}^* \rceil _f
\label{usface}
\end{split}
\end{equation}
\begin{equation}
\begin{split}
\lceil \bm u_{g} \rceil _f =
\left(\bm u^{pre}_{g}\right)_f + \frac{\lambda}{\Delta t}\left(\frac{\alpha_g \rho_g}{\mathbb D_g}\right)_f \left[\lceil \bm u^i_{g} \rceil _f - \left( \bm u^i_g \right)_f \right]
+ (1 -\lambda)\left[\lceil \bm u^*_{g} \rceil _f
    - \left(\bm u^*_g \right)_f \right]
    - \left(\frac{\alpha_g}{\mathbb D_g}\right)_f \nabla^{\perp}p_g + \left(\frac{\alpha_g\rho_g}{\mathbb D_g}\right)_f \bm g
    +\left(\frac{\beta}{\mathbb D_g}\right)_f \lceil \bm u_{s}^* \rceil _f
\label{ugface}
\end{split}
\end{equation}
where the phase pseudo-velocity is defined as $\bm u^{pre}_{k}=\frac{\mathbb H_k}{\mathbb D_k}+\frac{\lambda}{\Delta t}\frac{\alpha_k \rho_k}{\mathbb D_k} \bm u^i_{k}+(1 -\lambda)\bm u^*_{k} $.
Thus, a general momentum interpolation form that is independent of the time step size, transient discretization scheme and under-relaxation factor for arbitrary-shaped mesh is obtained. Eqs. (\ref{usface},\ref{ugface}) are the first complete momentum interpolation method for Euler-Euler simulation of gas-solid flows, to the best of our knowledge.
The corrections corresponding to the transient term $\frac{\lambda}{\Delta t}\left(\frac{\alpha_k \rho_k}{\mathbb D_k}\right)_f \left[\lceil \bm u^i_{k} \rceil _f - \left( \bm u^i_k \right)_f \right]$ and the under-relaxation term $(1 -\lambda)\left[\lceil \bm u^*_{k} \rceil _f
    - \left(\bm u^*_k \right)_f \right]$ are absent in previous studies on gas-solid two-phase flows \citep{passalacqua2011ImplementationIterativeSolution,liu2014cfd,venier2016numerical}. Moreover, the transient correction term exists in the current OpenFOAM$^\circledR$ version of the standard solver for gas-solid two-phase flow.
Eqs. (\ref{usface}, \ref{ugface}) are also the momentum interpolation method used in the validation cases in this work.

Additionally, the phase pseudo-velocity values $\bm u^{pre}_{k}$ can be obtained by explicitly treating the term $\mathbb H_k$ and solving simple linear algebraic equations in present study, thus avoiding the iterative solution of the nonlinear momentum equations and saving the computational cost \citep{rusche2003computational}. The absence of a step to solve the momentum equation iteratively allows us to set $\lambda=1$ so that the correction for under-relaxation can be discarded \citep{kobayashi1991numerical}. Therefore, the under-relaxation correction term is absent in the current OpenFOAM$^\circledR$ version of the standard solver for gas-solid two-phase flow, which means the momentum interpolation method used in present study and in the standard OpenFOAM$^\circledR$ solver is exactly same.

\section{Gas pressure Poisson equation and solid-phase continuum equation}
The gas pressure Poisson equation and solid-phase continuum equation are exactly same as in many previous studies  \citep{passalacqua2011ImplementationIterativeSolution,venier2016numerical,liu2024ps}, but with the generalized momentum interpolation method presented in previous section (Eqs. (\ref{usface}, \ref{ugface})) for calculating the volumetric momentum flux at cell faces. The governing equation for updating the gas pressure field is \citep{passalacqua2011ImplementationIterativeSolution}
    \begin{equation}
    \sum_f \left \{
                \left[
        (\alpha_{s})_f\left(\frac{\alpha_s}{\mathbb D_s}\right)_f + (\alpha_{g})_f \left(\frac{\alpha_g}{\mathbb D_g }\right)_f
    \right]
    |\bm S|\nabla^{\perp} p_g
        \right\}
    = \sum_f \varphi^0,
    \label{Poisson}
    \end{equation}
    where $\sum_f$ denotes the sum of all faces of the current cell, $\varphi^0$ is the total volumetric flux without the contribution of the gas pressure gradient term, it is expressed as
    \begin{equation}
    \varphi^0 =  (\alpha_{s})_f \varphi^0_s + (\alpha_{g})_f \varphi^0_g,
    \label{phi0}
    \end{equation}
    with
    \begin{equation}
    \varphi^0_s = \lceil \bm u_s \rceil _f \cdot \bm S + \left(\frac{\alpha_s}{\mathbb D_s} \right)_f\nabla^{\perp}p_g|\bm S|,
    \label{phic0}
    \end{equation}
    \begin{equation}
    \varphi^0_g = \lceil \bm u_g \rceil _f \cdot \bm S + \left(\frac{\alpha_g}{\mathbb D_g} \right)_f\nabla^{\perp}p_g|\bm S|.
    \label{phif0}
    \end{equation}

    The semi-discretized solid phase continuity equation is \citep{passalacqua2011ImplementationIterativeSolution}
    \begin{equation}
\frac{\partial \alpha_s}{\partial t} + \sum_f [(\alpha_s)_f \varphi_s'] - \sum_f \left[(\alpha_s)_f \left(\frac{1}{\mathbb D_s \rho_s}\frac{\partial p_s}{\partial \alpha_s}\right)_f |\bm S| \nabla^{\perp} \alpha_s \right] = 0,
    \label{phif2}
    \end{equation}
    where
    \begin{equation}
\varphi_s'=\varphi_s + \left( \frac{1}{\mathbb D_s \rho_s} \frac{\partial p_s}{\partial \alpha_s}\right)_f |\bm S| \nabla^\perp \alpha_s.
    \label{phif3}
    \end{equation}
After the solution of solid phase continuity equation, the gas volume fraction is updated as $\alpha_g=1-\alpha_s$.

\section{Flux reconstruction method for updating velocity fields}
The collocated variable arrangement combined with Rhie-Chow correction \citep{rhie1983numerical} modifies the gas pressure Poisson equation and incorporates numerical diffusion to attain a smoother and non-oscillatory gas pressure field. By doing so, it is possible to eliminate the high-frequency oscillations in the gas pressure field within the momentum equation. However, it is equally important to address this issue during the process of updating the velocity field. Otherwise, the efforts made through the momentum interpolation method will be nullified. Once the gas pressure equation (Eq. (\ref{Poisson})) is solved, the new gas pressure field stored at the cell center can be obtained and should be used to correct the volumetric flux of each phase stored at the face center.
After that, when updating the $k$-phase velocity fields at the cell center, the partial reconstruction method is used in order to eliminate the high-frequency oscillations of the gas pressure, gravity force and the granular pressure in the standard solver of the latest version of OpenFOAM$^\circledR$ \citep{weller2002derivation,rusche2003computational}:
    \begin{equation}
    \bm u_s = \bm u^{pre}_s + \frac{\beta}{\mathbb D_s}\bm u^*_{g} + \mathcal R_0\left [-\left(\frac{1}{\mathbb D_s}\right)_f \nabla^{\perp} p_s |\bm S| - \left(\frac{\alpha_s}{\mathbb D_s}\right)_f \nabla^{\perp} p_g |\bm S| + \left(\frac{\alpha_{s}\rho_s}{\mathbb D_s}\right)_f \bm g \cdot \bm S \right ],
    \label{usnew}
    \end{equation}
    \begin{equation}
    \bm u_g = \bm u^{pre}_g + \frac{\beta}{\mathbb D_g}\bm u^*_{s} + \mathcal R_0\left[- \left(\frac{\alpha_g}{\mathbb D_g}\right)_f \nabla^{\perp} p_g |\bm S|  + \left(\frac{\alpha_{g}\rho_g}{\mathbb D_g}\right)_f \bm g\cdot \bm S  \right],
    \label{ugnew}
    \end{equation}
    where $\mathcal R$ is a linear operator representing the process of reconstruction from the face fluxes into velocities, which attempts to preserve the smoothness, to provide a certain level of accuracy, and to yield cell-centered fields that are free from oscillations \citep{weller2014curl,weller2014non}. The subscript $0$ presents zero-order Taylor expansion:
    \begin{equation}
     \mathcal R_0(\varphi) = \left( \sum_f \frac{\bm S \bm S}{|\bm S|} \right)^{-1} \cdot \left[ \sum_f \varphi \frac{\bm S}{|\bm S|}\right],
     \label{recon0}
    \end{equation}
where $\varphi$ is a scalar field.
From Eqs. (\ref{usnew}, \ref{ugnew}), it can be seen that $k$-phase velocity $\bm u_{k}$ can be represented as a function of $\frac{\mathbb H_k}{\mathbb D_k}$, $\frac{\beta}{\mathbb D_{k}}\bm u_{1-k}$, $\mathcal R_0 \left[- \left(\frac{\alpha_k}{\mathbb D_k}\right)_f \nabla^{\perp} p_g |\bm S|  + \left(\frac{\alpha_{k}\rho_k}{\mathbb D_k}\right)_f \bm g\cdot \bm S  \right]$, and $\mathcal R_0 \left[-\left(\frac{1}{\mathbb D_s}\right)_f \nabla^{\perp} p_s |\bm S|\right]$ for solid phase. This suggests that although the high-frequency oscillations of gas  pressure field are suppressed through the use of the Rhie-Chow interpolation and the reconstruction, the updated velocity fields remain susceptible to the high-frequency oscillation of other cell-centered terms as $\frac{\mathbb H_k}{\mathbb D_k}$ and $\frac{\beta}{\mathbb D_{k}}\bm u_{1-k}$. In addition, the updated and smoothed gas pressure gradients in Eqs. (\ref{usnew}, \ref{ugnew}) cannot mitigate these issues as $\frac{\mathbb H_k}{\mathbb D_k}$ and $\frac{\beta}{\mathbb D_{k}}\bm u_{1-k}$ do not appear in the pressure Poisson equation due to the interpolation process. As a consequence, there is no feedback between the gas pressure gradient and the oscillations of phase velocities \citep{aguerre2018oscillation}.
Furthermore, the study of single-phase flow by \cite{aguerre2018oscillation} has shown that the zero-order flux reconstruction will degenerate the accuracy of CFD simulations.

Therefore, we propose that the $k$-phase velocity in Euler-Euler two-fluid model is calculated through certain kind of linear reconstruction operator $\mathcal R$ using $\varphi_s$ and $\varphi_g$:
    \begin{equation}
    \bm u_k^{i+1} = \mathcal R_1(\varphi_k) = \left(\sum_f \frac{\bm S \bm S}{|\bm S|}\right)^{-1} \cdot  \left\{ \sum_f \left[ \varphi_k - (\nabla \bm u_k^i)_f \cdot \bm d_{Pf} \cdot \bm S \right] \frac{\bm S}{|\bm S|}  \right\},
    \label{recon2}
    \end{equation}
    where the superscript $i$ represents the value after previous iteration convergence. The first-order Taylor expansion is applied here instead of the zero-order Taylor expansion to achieve higher accuracy. Eq. (\ref{recon2}) presents a novel reconstruction form that is originally proposed by \citet{aguerre2018oscillation} for single-phase flow so that the velocity field that is free of oscillations can be obtained.
    Consequently, the primitively update step of the cell-centered phase velocity fields that are available in the latest OpenFOAM$^\circledR$ solver (Eqs. (\ref{usnew},\ref{ugnew})) should be replaced by Eq. (\ref{SS_phi}):
    \begin{equation}
    \begin{split}
    \bm u_k = &\mathcal R_1\Bigg \{ \bm u^{pre}_k \cdot \bm S + \frac{\lambda}{\Delta t}\left(\frac{\alpha_k \rho_k}{\mathbb D_k}\right)_f\left[\varphi^i_{k} - (\bm u^i_k)_f \cdot \bm S \right]
    + (1 -\lambda)\left[ \varphi_{k}^* - \left(\bm u_k^*\right)_f \cdot \bm S \right]
    +\left(\frac{\beta}{\mathbb D_k}\right)_f \varphi_{1-k}^* \\
    & - \zeta \left(\frac{1}{\mathbb D_s}\right)_f\nabla^{\perp}p_s |\bm S|
    -\left(\frac{\alpha_k}{\mathbb D_k}\right)_f \nabla^{\perp}p_g |\bm S|+ \left(\frac{\alpha_k \rho_k}{\mathbb D_k}\right)_f \bm g \cdot \bm S \Bigg \}.
    \end{split}
    \label{SS_phi}
    \end{equation}

With the present flux reconstruction method, the gas and solid phase velocities are updated by their complete volumetric fluxes at the interfaces that are obtained by suitable momentum interpolation method, so that the high-order oscillations can be better suppressed.

    \section{Validation cases}
    In the present investigation, a comprehensive computational analysis is conducted to simulate a multitude of scenarios, encompassing a particle settling, a fixed bed case, three distinct cases of fluidized beds characterized by varying superficial gas velocities and two cases of circulating fluidized beds characterized by varying solid fluxes. The axial distribution of the cross-sectional average parameters will serve as the foundation for the analytical descriptions. Therefore, this study primarily examines the discrepancy between the numerically obtained values and the theoretically accurate values, other CFD results are not reported for the purpose of simplicity. Euler-Euler simulations of gas-solid flows are carried out using the latest version of OpenFOAM$^\circledR$ solver (denoted as Standard Solver hereafter), where the center-to-face momentum interpolation method of Eqs. (\ref{usface},\ref{ugface}) is used and the face-to-center flux reconstruction is performed using  Eqs. (\ref{usnew}, \ref{ugnew}), and an OpenFOAM$^\circledR$ solver with Eqs. (\ref{usface},\ref{ugface}) for the center-to-face momentum interpolation and Eq. (\ref{SS_phi}) for the face-to-center flux reconstruction (Denoted as This Work hereafter). It can be seen that the momentum interpolation method of the standard solver and the solver in this work remain consistent, the only difference is the flux reconstruction method. In addition, all other models and algorithms are exactly the same.

        \subsection{Solids settling}
        \begin{figure}[!htb]
        \centering
        \includegraphics[scale=0.5]{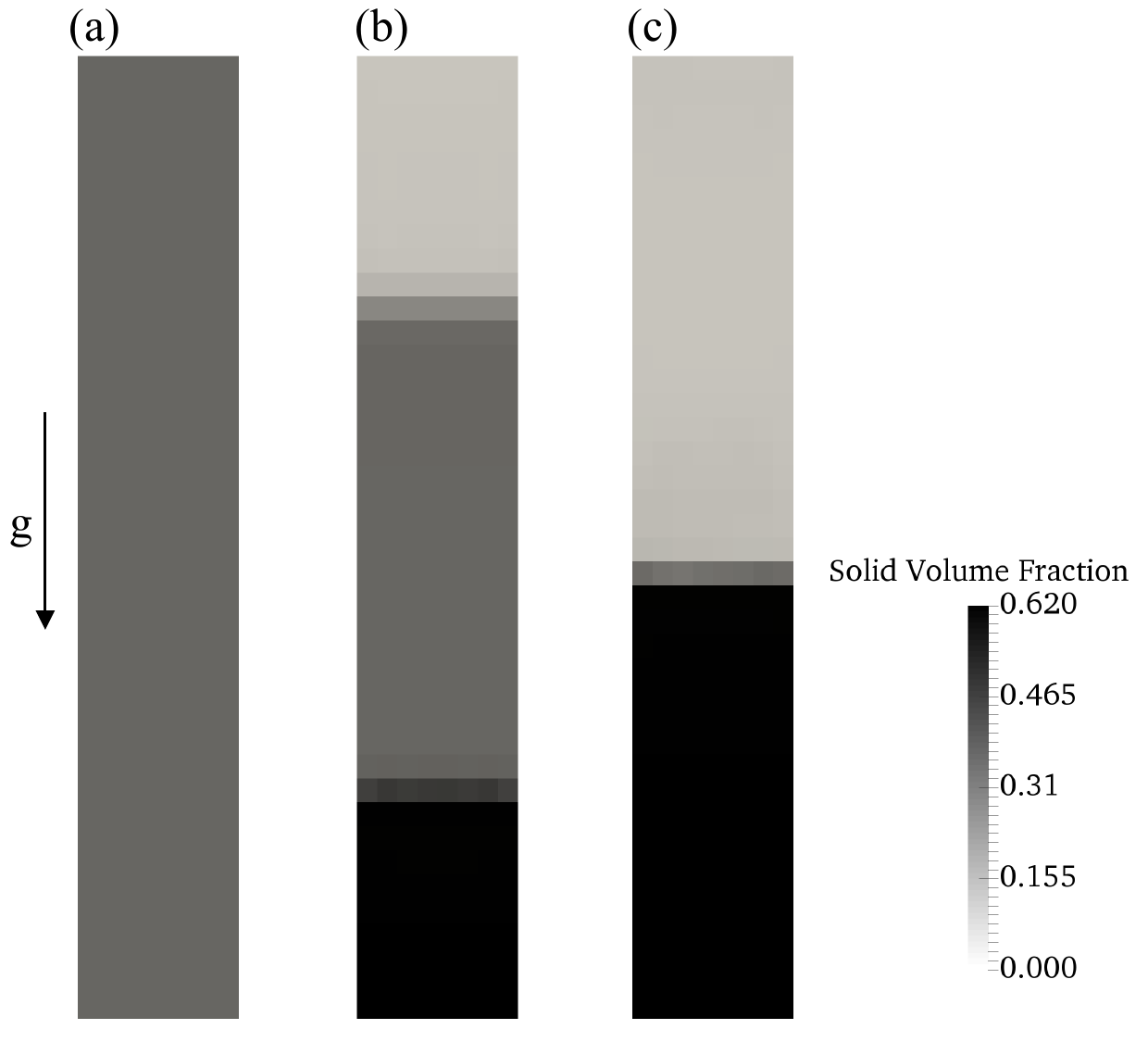}
        \caption{Snapshots of the solid volume fraction at different times during solids settling process: (a) The initial state, where the particles are evenly distributed throughout the entire bed with a given volume fraction of 0.3; (b) The intermediate state, the upper part consists of pure gas while the lower part is composed of particles that have reached the packing limit. The intermediate region is a mixture of the two; (c) The stable state of complete settlement is reached, with a clear demarcation between the upper part consisting of pure gas and the lower part consisting of closely packed particles.}
        \label{solidssettling_contour}
        \end{figure}

        \begin{figure}[!htb]
        \centering
        \includegraphics[scale=0.5]{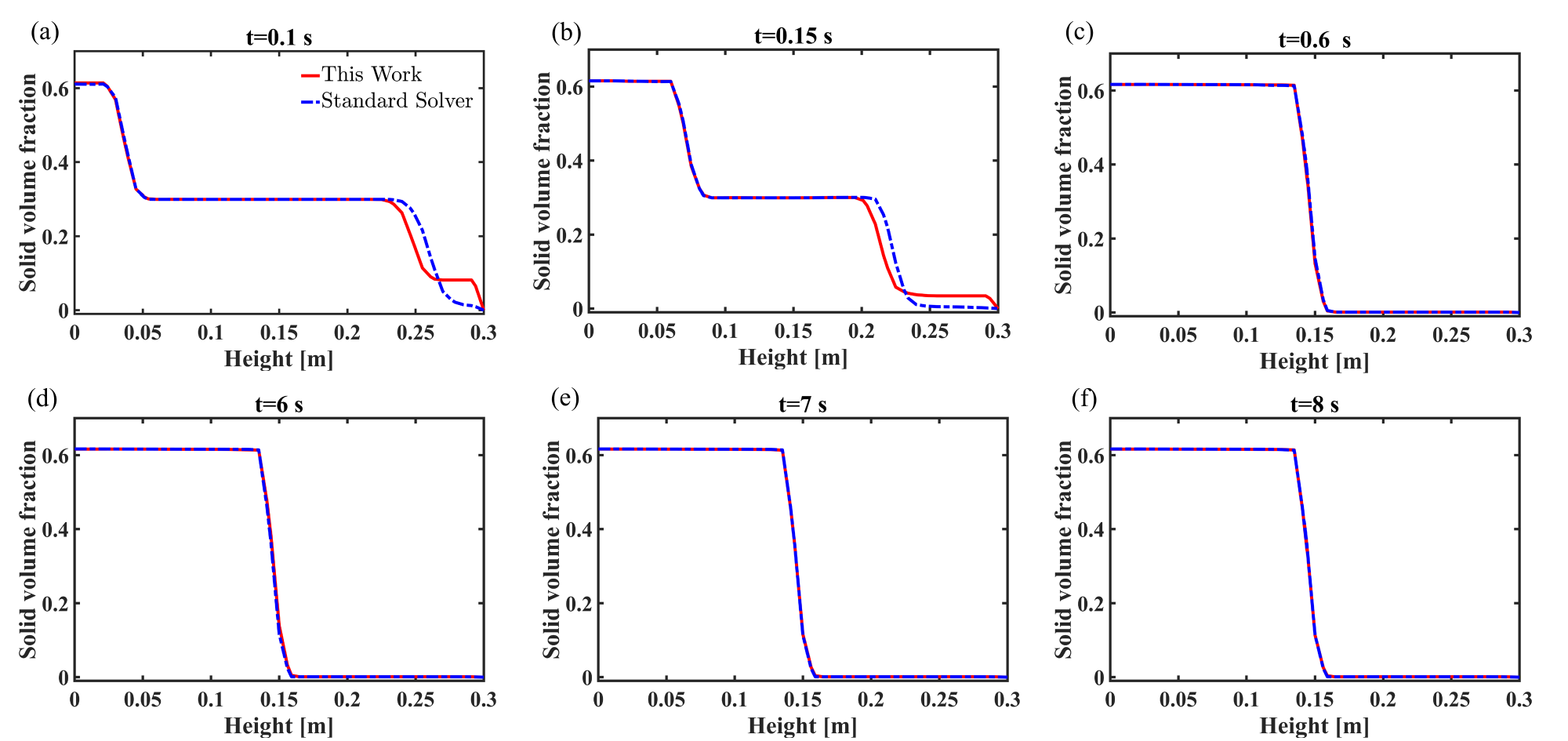}
        \caption{Solid volume fraction distribution along the tube height using different flux reconstruction methods at (a) t=0.1 s, (b) t=0.15 s, (c) t=0.6 s, (d) t=6 s, (e) t=7 s, (f) t=8 s.}
        \label{solidssettling_es}
        \end{figure}

        \begin{figure}[!htb]
        \centering
        \includegraphics[scale=0.8]{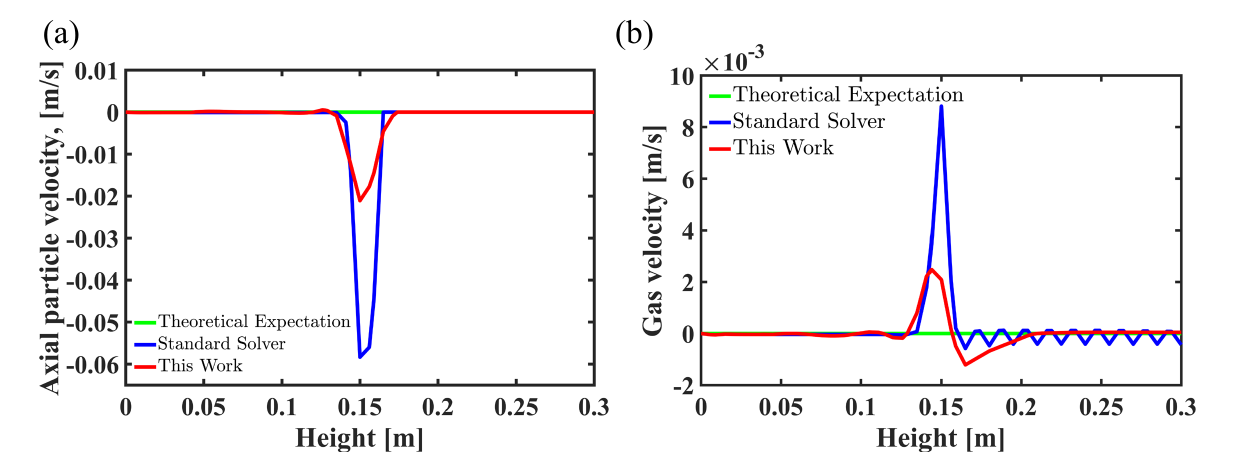}
        \caption{Time-averaged axial (a): solid velocity ($\bm u_{s,y}$) and (b): superficial gas velocity ($\bm U_{g,y}$) distribution along the vertical height using different flux reconstruction methods.}
        \label{solidssettling_u}
        \end{figure}

The verification example is composed of a two-dimensional vertical bed. Uniformly distributed particles settle down to the bottom due to the gravity. The settling process is illustrated in Fig. \ref{solidssettling_contour}. The initial solid volume fraction is 0.3. The bed has a height of 0.3 m and a diameter of 0.05 m, discretized by 40 $\times$ 8 orthogonal uniform grids. The gas density is 1.2 kg/m$^3$ with a dynamic viscosity of 1.8e-5 Pa s. The particle density is 2000 kg/m$^3$ and particle diameter is 4e-4 m. The restitution coefficient is 0.6. The Gidaspow drag model is adopted as shown in Table \ref{tab:drag} in Appendix A. The Schaeffer model is used for closing frictional pressure and the kinetic theory of granular flow is used to close granular pressure as shown in Table \ref{tab:Constitutive} in Appendix A.
There is no particle and gas in or out. Particles and gas are set to free slip and no-slip at the walls, respectively. The pressure is set to the atmosphere at the outlet and zero-gradient at inlet and walls.
The transient term is discretized by the second-order Euler implicit scheme, the convective term is discretized by the Gauss limited linear scheme, which is a kind of TVD scheme. The gradient term is discretized by the central difference scheme. The maximum residual for the pressure and continuity equations are 1e-8 and 1e-9, respectively. Each time step consists of three cycles of the solid phase continuity equation and three cycles of the PIMPLE iteration, all of which satisfy the convergence criterion.
The simulation is carried out for 8 s totally with a time step of 1 $\times$ 10$^{-4}$ s and the time-averaged statistics were conducted from 6 s to 8 s.

Fig. \ref{solidssettling_es} presents the transient solid volume fraction distribution along the vertical line of the tube for two different flux reconstruction methods. It can be seen that the steady state has been reached at t = 0.6 s. After reaching the steady state, both solvers show a consistent axial distribution of solid volume fraction.
Fig. \ref{solidssettling_u}a is the time-averaged profile of particle velocity component in the axial direction ($\bm u_{s,y}$) along the vertical lineusing different flux reconstruction methods. Since there is no gas entering, particles are settled due to gravity. Once the system reaches the steady state, the entire bed is stationary and the particle velocity should be zero throughout the bed. In the uppermost region of the single-gas phase flow, both employed methodologies yield a velocity profile that approaches zero precisely. In addition, within the lower region of the bed where the packing limit is attained by particles, the velocity profile approaches zero as well, as evidenced by both methods. This test case is particularly intricate due to the substantial transition in the solid volume fraction at the interface of these two zones, which can be accurately predicted by both approaches. However, the standard solver exhibits higher levels of spurious velocities due to the abrupt change of state variables, the issue is better mitigated by the employment of the novel flux reconstruction technique, although spurious velocities still exist.
Fig. \ref{solidssettling_u}b is the time-averaged profile of gas velocity component in the axial direction ($\bm U_{g,y}$) along the vertical line using different flux reconstruction methods. Spurious velocity values of gas velocity also appear at the interface and are significantly suppressed after using the new flux interpolation method. Additionally, in the upper single-phase flow region, the results of the standard solver exhibit high-frequency oscillations around the expected value, which can be filtered out by the new flux interpolation method.

\subsection{Fixed bed}
A two-dimensional gas-solid fixed bed is simulated in this work. The fixed bed has a height of 1.0 m and a diameter of 0.28 m, discretized by 56 $\times$ 200 grids. The experimentally determined minimum fluidization velocity is $\bm U_{mf} = 0.065m/s$. Initially, the static bed height is 0.4 m and the solid volume fraction is 0.6, with a particle diameter of $d_p = 275 \mu m$, a particle density of $\rho_s = 2500 kg/m^3$, and a restitution coefficient of $e=0.9$. The gas density is $\rho_g = 1.225kg/m^3$, with a viscosity of $\mu_g = 1.485 \times 10^{-5} kg/m \cdot s$. At the bottom inlet, uniform superficial gas velocity of 0.03 m/s is defined without particles. At the wall, no-slip condition is employed for gas, Johnson-Jackson condition is employed for particles and zero-gradient condition is employed for pressure. At the top outlet, the pressure is fixed to the atmosphere, i.e. 101325 Pa. The transient term is discretized by the first-order Euler implicit scheme, the convective term is discretized by the second-order TVD scheme (Gauss linear 1), and the diffusion term is discretized by the central difference scheme. The under-relaxation factor is 0.3, 0.2 and 0.2 for pressure equation, solid-phase continuity equation and granular equation, separately. The Johnson-Jackson Model (Table \ref{tab:Constitutive}) is used to close frictional pressure and the kinetic theory of granular flow is applied to close granular pressure. The Syamlal-O'Brien drag model (Table \ref{tab:drag}) is applied. A total time of 30 s were conducted with a time step of 1 $\times$ 10$^{-4}$ s and the data of the last 5 s were counted as the average result.

        \begin{figure}[!htb]
        \centering
        \includegraphics[scale=0.4]{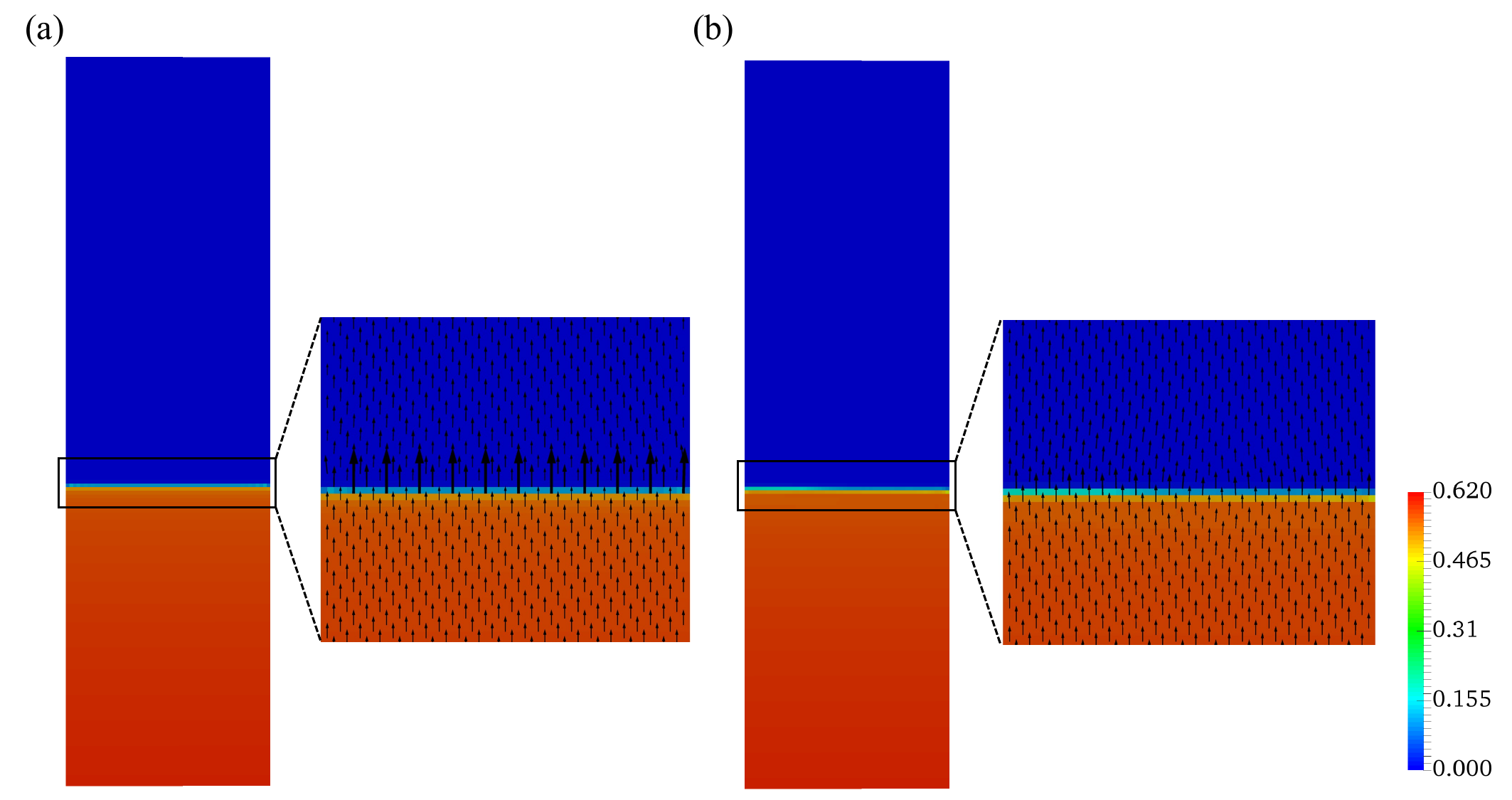}
        \caption{Time-averaged solids concentration contours with the superficial gas velocity vectors of (a): standard solver, (b): this work.}
        \label{contour}
        \end{figure}

        \begin{figure}[!htb]
        \centering
        \includegraphics[scale=0.8]{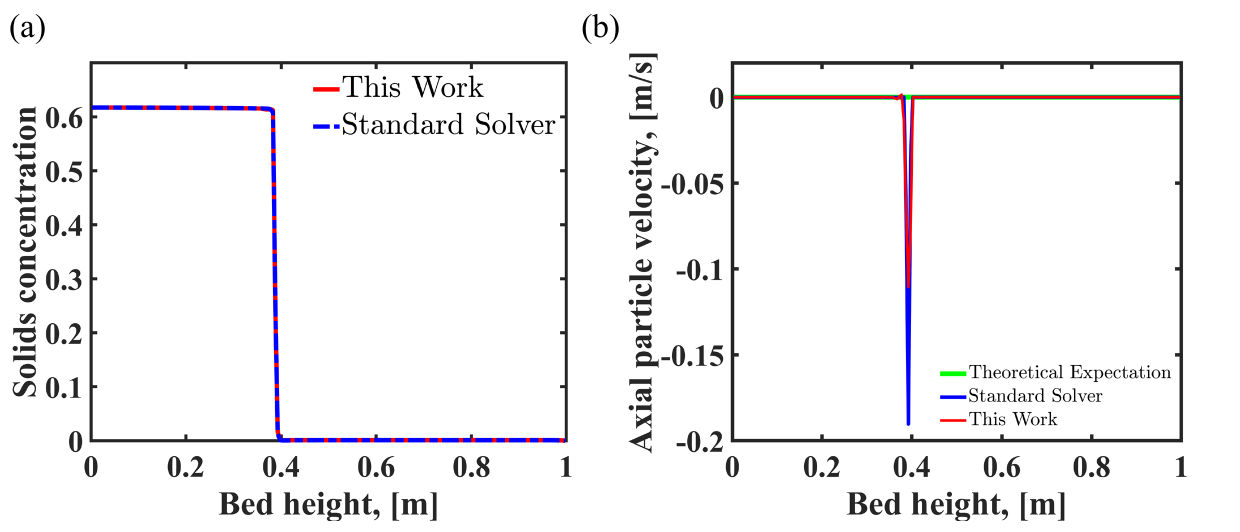}
        \caption{Time and cross-sectional averaged (a): solids concentration ($\alpha_s$) and (b): axial particle velocity ($\bm u_s$) profiles using different flux reconstruction methods.}
        \label{fixedbed}
        \end{figure}

        \begin{figure}[!htb]
        \centering
        \includegraphics[scale=0.5]{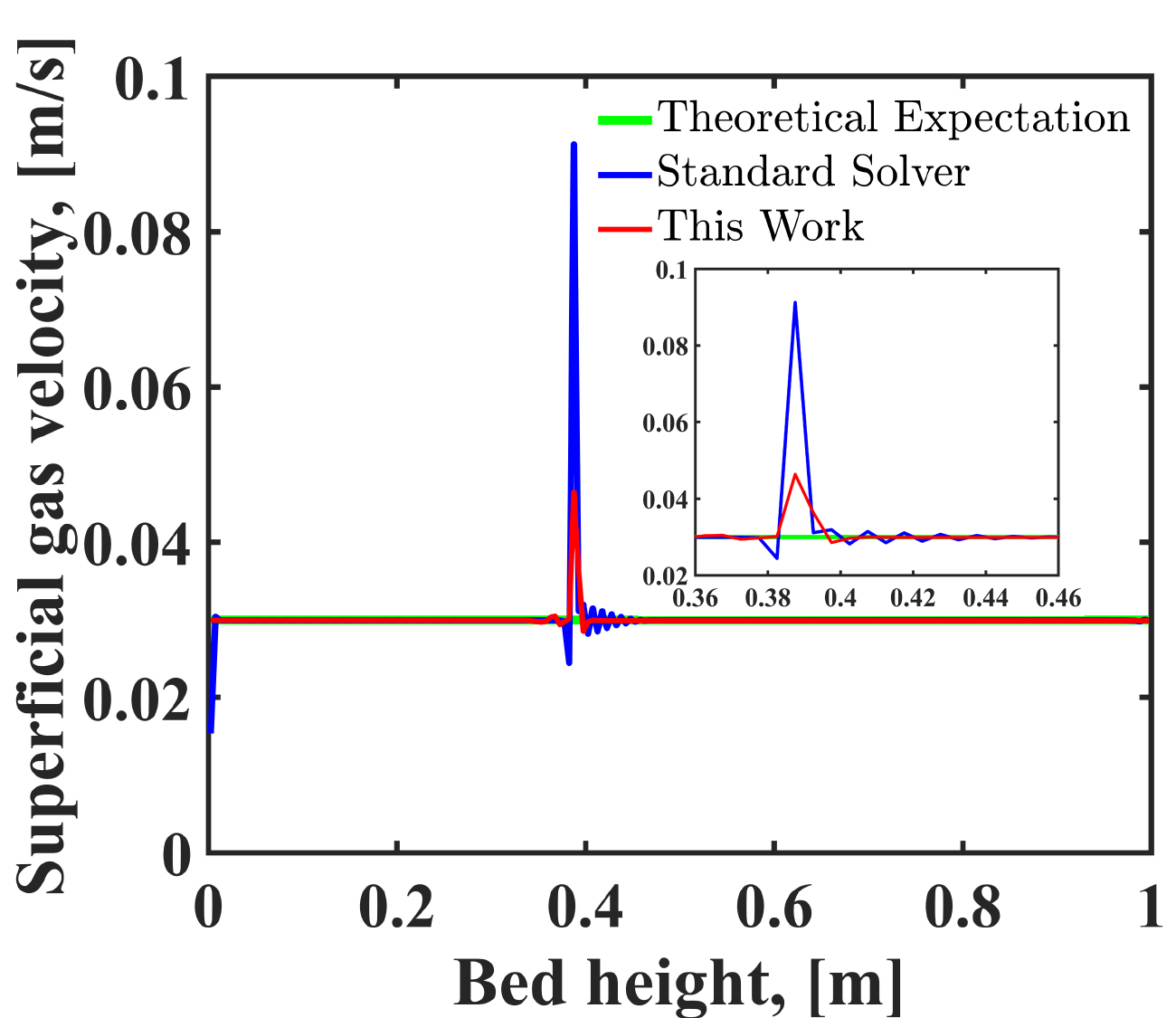}
        \caption{Time and cross-sectional averaged axial superficial gas velocity ($\bm U_g$) profiles using different flux reconstruction methods.}
        \label{fixedbed_Ugy}
        \end{figure}

The time-averaged solids concentration contours obtained by the two methods are shown in Fig. \ref{contour}, accompanied by the superficial gas velocity vector. The lower section is characterized by a tightly packed particles, whereas the upper section is the single-phase gas flow. A distinct interface exists between them, marking a region of substantial gradient where the flow dynamics are susceptible to transitionary phenomena. At the interface, the superficial gas velocity vector predicted by the standard solver is considerably larger than that predicted by the new reconstruction method, which is more uniform.

For the fixed bed regime studied in this section, the inflow superficial gas velocity of 0.03 m/s is less than the minimum fluidization velocity of 0.065 m/s, which is determined experimentally \citep{taghipour2005experimental}. Within the bed, the theoretical expectation of particle velocity should be 0 m/s, as in the case of solids settling. Furthermore, in accordance with the fundamental tenets of the finite volume method, if we consider a cross section as a control volume, in the absence of a mass source, the mass of gas entering the unit cross-section must be qual to the mass of gas leaving this cross-section: $(\alpha_g \rho_g \bm u_g)$, where the subscript $i$ denotes the number of mesh layers along the mainstream direction. For incompressible fluids, the density is constant and the definition of the superficial velocity $\bm U_g = \alpha_g \bm u_g$, thus we have $(\bm U_g)_{inlet} = (\bm U_g)_{i}$, which means the per unit cross-sectional averaged superficial gas velocity should be equal to $\bm U_{g,inlet}$.

Fig. \ref{fixedbed} presents the time and cross-sectional averaged axial solids concentration and particle velocity profiles using different flux reconstruction methods. Spurious particle velocities exist only at the interface and are significantly suppressed when using the new reconstruction method. Fig. \ref{fixedbed_Ugy} presents the time and cross-sectional averaged axial superficial gas velocity profiles with its enlargement at the interface. The standard solver predicts spurious superficial gas velocities far from the expected value at the interface, while the new flux reconstruction method can reduce these deviations significantly. It is also obvious that there are high-frequency oscillations at the interface described by the standard solver and can be filtered out by the new reconstruction method. However, it is not yet possible to eliminate them entirely due to the large solids concentration gradient at the interface under the circumstance of the current grid resolution and discretization scheme. Furthermore, at the inlet, the standard solver also produces values that are implausible and deviate significantly from the expected value. This may be attributed to the bias introduced by the momentum transport term during the flux reconstruction process of velocity. Following the implementation of the novel flux reconstruction methodology, the observed anomalous behaviour at the inlet was successfully mitigated.\\

\subsection{Bubbling fluidized bed}
A two-dimensional gas-solid fluidized bed with Geldart B particles on the experimental basis of \citet{taghipour2005experimental} is simulated. The same case has also been numerically investigated by \citet{taghipour2005experimental,herzog2012comparative}. The fluidized bed has the same height, diameter and mesh resolutions as the fixed bed. Additionally, the initial conditions, boundary conditions, the physical characteristics, the numerical schemes, the drag model and the granular pressure models are also consistent with those in the fixed bed. Three cases with varying superficial gas velocities of 0.2, 0.38, 0.46 m/s are investigated in this study. We however only report the result of 0.2 m/s, since the results of another two superficial gas velocities are qualitatively same. A total time of 30 s were conducted with a time step of 1 $\times$ 10$^{-4}$ s and the data of the last 5 s were counted as the averaged results.

Increasing the gas velocity leads to a continuous increase in the bed pressure drop until it equals the gravity per unit cross-sectional area of the bed. When the drag force exerted by the gas flow counterbalances the gravity of the particles, the particles start to be fluidized. Although in the bubbling fluidization regime the particles subject extensive motions and complex bubbling structures formed and evolved, the cross-sectional averaged mass fluxes of solids and gas are still conserved. Thus we have $(\bm U_s)_{inlet} = (\bm U_s)_{i}$ and $(\bm U_g)_{inlet} = (\bm U_g)_{i}$, therefore, the cross-sectional theoretical expectation for the particle velocity is 0 m/s, while the superficial gas velocities are 0.2, 0.38 and 0.46 m/s, respectively.

\begin{figure}[!htb]
        \centering
        \includegraphics[scale=0.65]{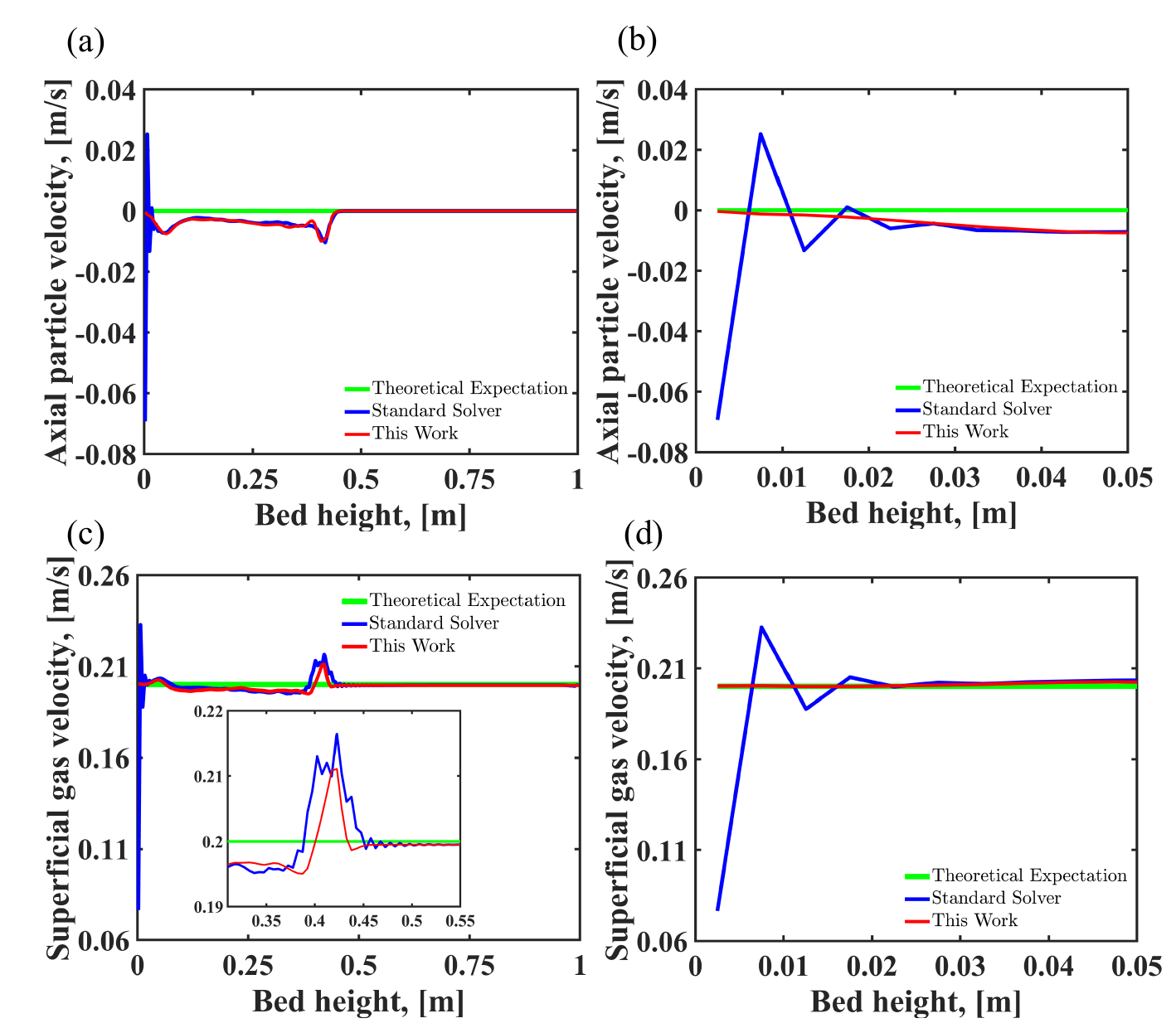}
        \caption{Time and cross-sectional averaged axial (a): particle velocity and (c): superficial gas velocity profiles using different flux reconstruction methods of $U_g = 0.2 m/s$ with their enlargements at the inlet region (b) and (d), respectively.}
        \label{bubbling_Axialus}
        \end{figure}

Fig. \ref{bubbling_Axialus} illustrates the time and cross-sectional averaged axial particle velocity and superficial gas velocity profiles using two flux reconstruction methods. Examination of the Fig. \ref{bubbling_Axialus} elucidate a close concordance with the theoretical expectations in the upper region of single-phase gas flow. Conversely, around the inlet region, the standard solver yielded gas and particle velocity data that exhibit pronounced oscillations whereas the proposed reconstruction method yielded a comparatively smooth velocity profiles (Figs. \ref{bubbling_Axialus} b,d). At the bed height location, both methods predict large spurious particle velocities at the bed height region (Figs. \ref{bubbling_Axialus} a). The standard solver was observed to furnish more oscillatory and less precise superficial gas velocity values relative to the proposed flux reconstruction method (Figs. \ref{bubbling_Axialus} c).

\subsection{Circulating fluidized bed riser}

        \begin{figure}
        \centering
        \includegraphics[scale=0.5]{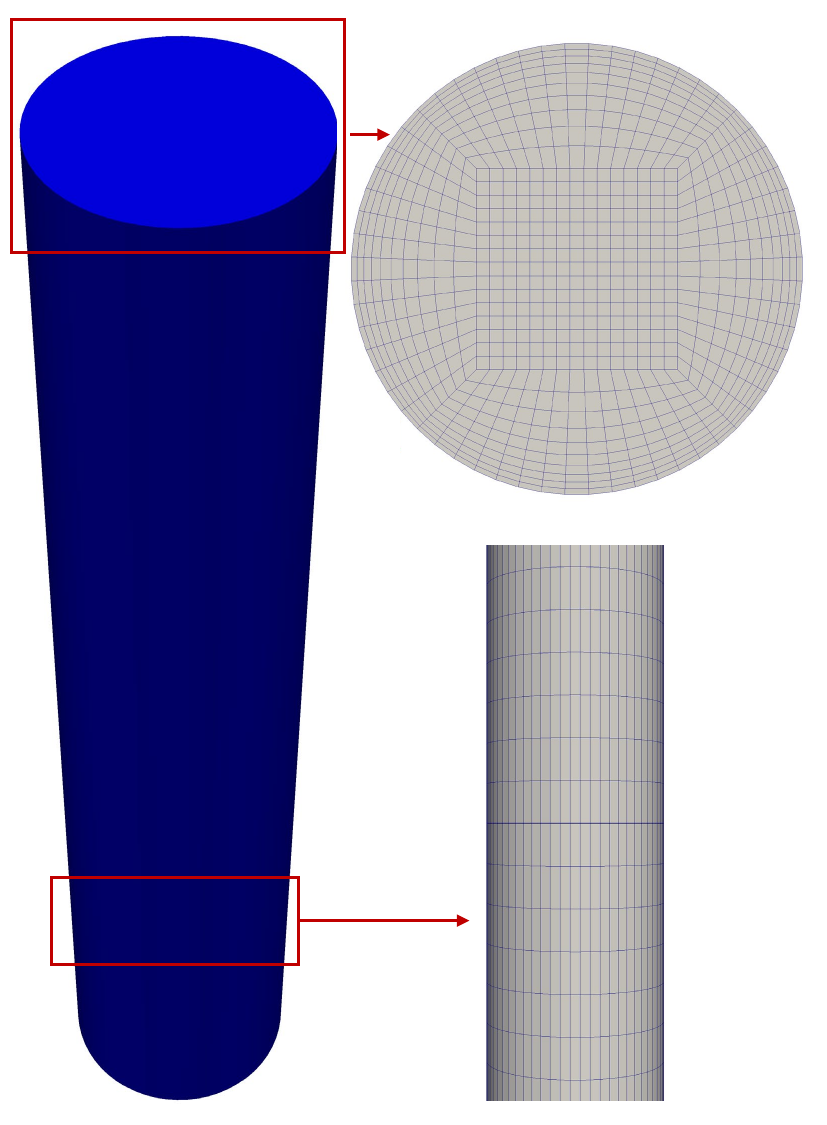}
        \caption{Schematic geometry and grid of the three-dimensional riser (For ease of presentation, the riser is not at scale).}
        \label{3Dgrids}
        \end{figure}

        \begin{figure}
        \centering
        \includegraphics[scale=0.5]{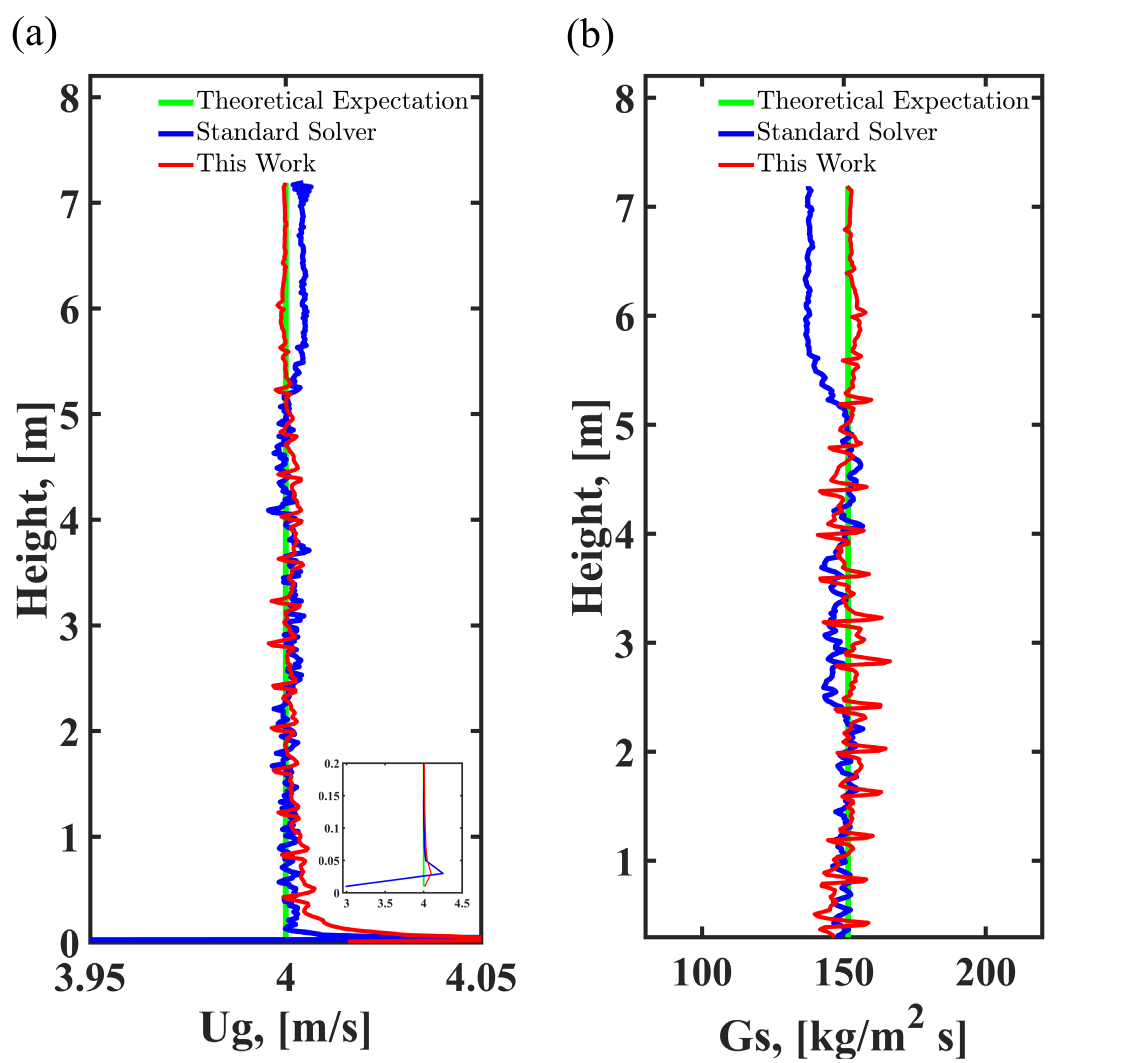}
        \caption{Time and cross-sectional averaged axial (a) superficial gas velocity ($U_g$) with its enlargement at the inlet and (b) solid flux profiles in the riser with a solid flux of $Gs=150.1kg/m^2s$.}
        \label{riser_AxialUg}
        \end{figure}

A three-dimensional circulating fluidized bed riser is simulated. As illustrated in Fig. \ref{3Dgrids}, the riser has a height of 7.2 m and a diameter of 0.09 m. The total number of grids is 297000, with the grid number of 360 in the mainstream direction. The superficial gas velocity at the bottom inlet is $\bm U_g$ = 4 m/s. The solid volume fraction is specified to be zero-gradient and the corresponding solid mass fluxes are $G_s$ = 138.5 kg/m$^2$ s and 150.1 kg/$m^2$ s, respectively. The outlet is situated at the top, where the atmospheric pressure is specified. The no-slip wall condition is imposed for the gas phase while the free slip wall condition is imposed for the solid phase. Initially, the averaged solid volume fraction in the whole riser was specified as $\alpha_{s,ini}$ = 0.055. The particle diameter is $d_p$ = 100 $\mu$ m. The particle density is $\rho_s$ = 2650 kg/m$^3$ with a restitution coefficient of e = 0.95. The gas density is $\rho_g$ = 1.2 kg/m$^3$, with a viscosity of $\mu_g$ = 1.8e-5 kg/m $\cdot$ s. The transient term is discretized by first-order Euler implicit scheme, the convective term is discretized by the second-order TVD scheme (Gauss linear 1). The diffusion term and the gradient term is discretized by the central difference scheme. The under-relaxation factor is 0.5, 0.2 and 0.2 for pressure equation, solid-phase continuity equation and granular energy equation, respectively. The frictional stress is closed by the Johnson-Jackson model and the granular pressure is closed by the kinetic theory of granular flow as shown in Table \ref{tab:Constitutive}. The EMMS-based drag model (Table \ref{tab:drag}) is applied to take the effect of particle clustering structures into consideration, nevertheless, particulars of the drag correlation for each simulation cases are not described in detail. The simulations are performed as transient for 60 s in total with the fixed time step of $5\times10^{-5}$  s. After the initialization and fully development of flow, time-averaged data are then collected from the final 20 s. Again, the results of the two cases are similar, therefore, only $G_s=$150.1 kg/m$^2$s are reported.

Fig. \ref{riser_AxialUg} presents the time and cross-sectional averaged axial superficial gas velocity and solid flux profiles.
The averaged superficial gas velocity computed by the standard solver exhibits high-frequency oscillations throughout the entire height and there are discrepancies between the obtained values and the expected values, which is particularly pronounced in the inlet region below 0.1 m (Fig. \ref{riser_AxialUg}a enlargement) and the upper part above 5 m. By contrast, the proposed flux reconstruction method effectively mitigates these oscillations and reduces the error. The relative error between the simulated values and the theoretically anticipated values are appreciably decreased by the application of the novel flux reconstruction technique.
As the height increases, the cross-sectional averaged solid flux predicted by the standard solver deviates more and more from the theoretical expectation, and the error of solid mass non-conservation is considerable higher than 5 m. In contrast, the new flux reconstruction method can obtain much more accurate velocities, which are closer to the theoretical expectation value, and have more regular oscillations.

\begin{figure}[!htb]
    \centering
    \includegraphics[scale=0.7]{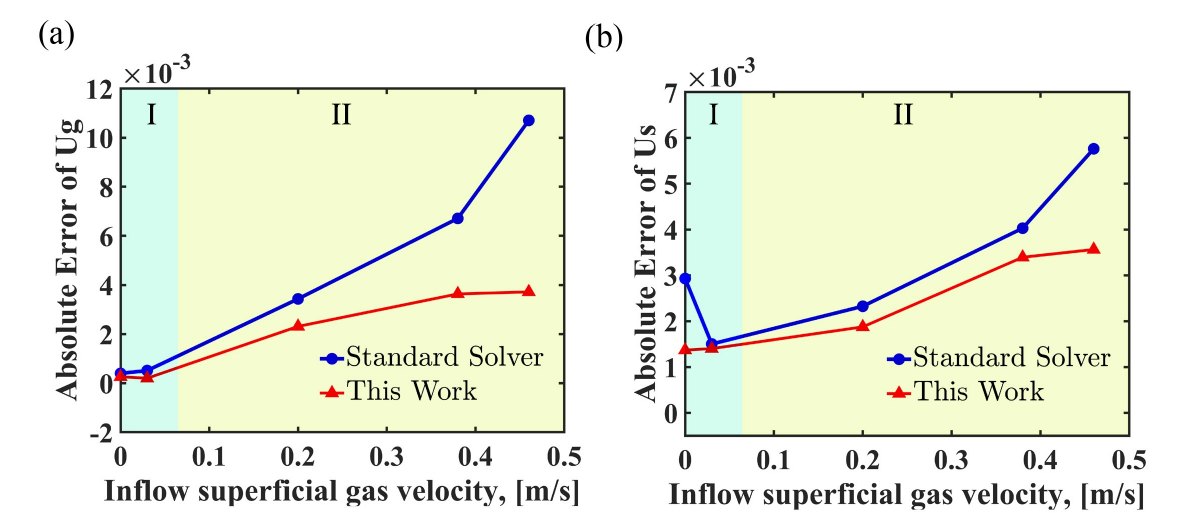}
    \caption{Absolute error between the numerical results and the theoretical expectations, (a): superficial gas velocity, (b): particle velocity, where $\uppercase\expandafter{\romannumeral1}$ is the fixed bed regime including the case of solids settling with the inflow superficial gas velocity $\bm (U_g)_{inlet}=0 m/s$ and the fixed bed with $\bm (U_g)_{inlet}=0.03m/s$, $\uppercase\expandafter{\romannumeral2}$ is the bubbling fluidized bed regime including cases with various inflow superficial gas velocities $\bm (U_g)_{inlet}=0.2, 0.38m/s$ and $0.46m/s$.}
    \label{error1}
\end{figure}

\begin{figure}[!htb]
    \centering
    \includegraphics[scale=0.6]{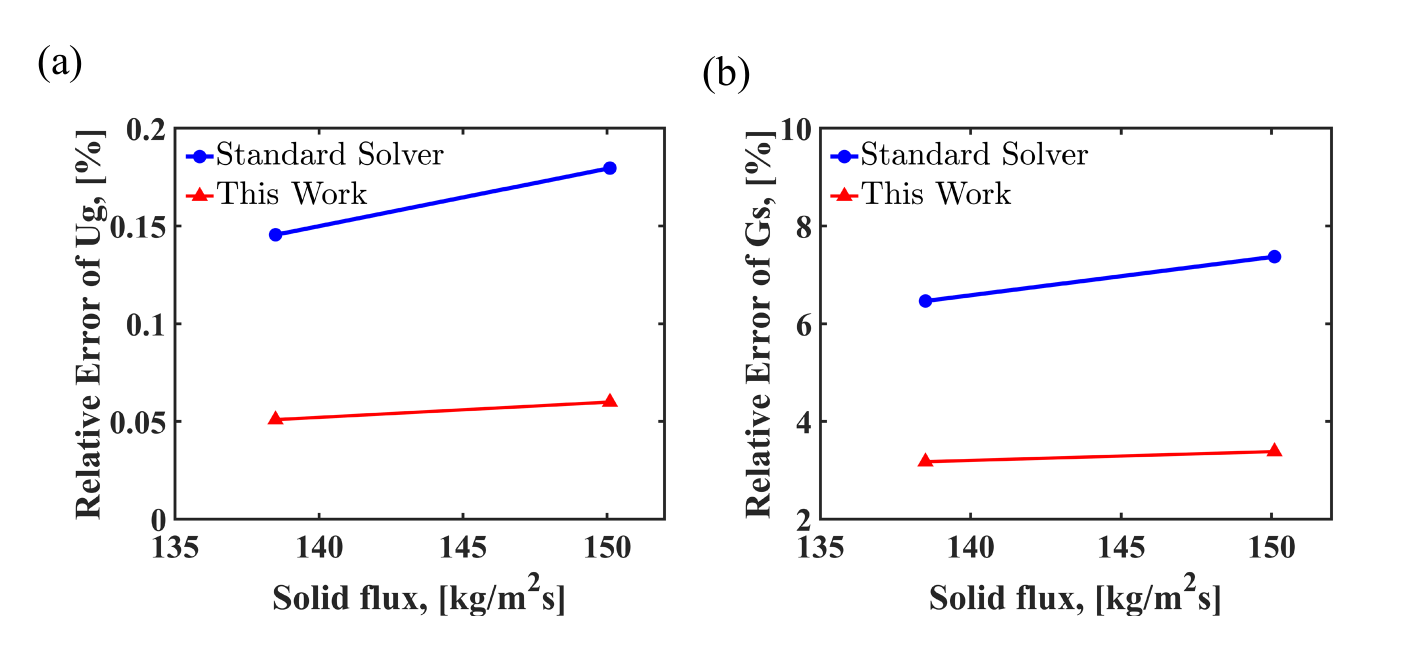}
    \caption{Relative error between the numerical results and the theoretical expectations, (a): superficial gas velocity, (b): solid flux in the riser including two cases with different solid fluxes $(G_s)_{inlet}=138.5$ and $150.1kg/m^2s$.}
    \label{error2}
\end{figure}

Since the theoretical expected value of the particle velocity in the solids settling, the fixed bed and the bubbling fluidized bed (BFB) cases are 0 m/s, the absolute error ($\frac1N |x-x_{expected}|$) are calculated for these cases, where $N$ is the number of data or the grid number in the mainstream direction. Fig. \ref{error1} illustrates that the value of absolute error increases from the fixed bed regime to BFB regime. In addition, for BFB cases, the absolute error value increases with the increase of the inflow superficial gas velocity.
While the relative error ($\frac1N \frac{|x-x_{expected}|}{x_{expected}}$) are calculated for the circulating fluidized bed (CFB) riser cases. Fig. \ref{error2} illustrates that the relative error value increases with the increase of the solid flux in CFB riser.
For the cases of fixed bed and bubbling fluidized beds that share the same phase physical parameters and geometric structures, as the inflow superficial gas velocity escalates, the degree of fluidization within the bed intensifies, resulting in an increase in bed height, the proliferation of bubbles, more gas-solid phase interfaces, and a concomitant enlargement of the region prone to velocity mutation. The same qualitative trend also occurs in the fast circulating bed riser along the direction of increasing solid flux. Consequently, the magnitude of velocity error escalates accordingly. However, the newly proposed flux reconstruction method exhibits a marked ability to mitigate these errors, enhancing the overall smoothness and precision of the velocity profiles across all simulated cases examined.
On the other side, it is also crystal clear that although the new flux reconstruction method can to a significant extent suppress the high-frequency oscillations, the velocity profiles do not conform entirely to the theoretical expectations, manifesting a degree of discrepancy. In the particle settling case and the fixed bed case, the error exists at the interface between the upper gas single-phase flow and the particles tightly packed region. In the BFB cases, in addition to the aforementioned interface, there are also errors within the bed. In the CFB cases, the error exists at the entire computational domain. All these regions have large gradients of the state variables.
Additionally, The preceding results demonstrate that as the spatiotemporal frequency of the gas-solid phase interface increases, the value of error rises. This is because in such circumstances, the frequency of the abrupt change in state variable increases.
Therefore, the observed discrepancy between the numerical solution obtained by this work and the theoretical expectation may be attributed to the inherent limitations of both the momentum interpolation method and the flux reconstruction method, which are formulated on the basis of Taylor series expansion. Such method is susceptible to inaccuracies when encountering abrupt changes in the independent variable, say, the edges of bubble-emulsion phases in bubbling fluidized beds and the interface between dilute and dense phases in CFB riser. 

\section{Conclusion}
In this study, in order to suppress the high-frequency oscillations of velocity, a novel flux reconstruction method for updating the cell-centered velocities through face flux that is first proposed to improve smoothness for single-phase flow \citep{aguerre2018oscillation} is extended to two-phase flow and combined with suitable momentum interpolation method, in which the zero-gradient Taylor series expansion in the standard solver of latest OpenFOAM$^\circledR$ version is replaced by the first-order expansion to enhance the precision. Therefore, the accuracy of both momentum transport terms and force balance are considered, which would suppress the high-frequency oscillation of the velocities to a certain extent and improve the accuracy. Several numerical cases are conducted and compared with the standard solver of OpenFOAM$^\circledR$ to verify the superiority of the new technique.
In the single-phase gas flow region and the region where the particles are closely packed, the gradient of the independent variable is zero, the velocity obtained by both methods has almost no high-frequency oscillation and the deviation caused by it is negligible. In contrast, in the two-phase flow region where large gradient of the independent variables exists, the velocity obtained by the standard solver exhibits oscillations and deviations from the theoretical expectation significantly. The new methodology is capable of effectively suppressing these oscillations and significantly reducing the deviations from the theoretical value.
However, the oscillations and deviations can not be eliminated completely through this work due to the inherent drawback of the momentum interpolation technique and flux reconstruction method, which are based on the Taylor series expansion. The methods in question are prone to inaccuracies in the presence of abrupt changes in the independent variables. Our future work will attempt to develop a satisfactory method for tackling this issue within the framework of Euler-Euler simulations of gas-solid flows.

\section*{Declaration of Competing Interest}
The authors declare that they have no known competing financial interests or personal relationships that could have appeared to influence the work reported in this paper.

\section*{Acknowledgement}
This study is financially supported by the National Natural Science Foundation of China (11988102, 22311530057), the Strategic Priority Research Program of the Chinese Academy of Sciences (XDA29040200), the Young Elite Scientists Sponsorship Program by CAST (2022QNRC001) and the Science Foundation of China University of Petroleum, Beijing (No.2462024YJRC008). We would like to thank Dr. Jinghong Su and Mr. Shaotong Fu for their valuable suggestions.

\section*{Appendix A: Constitutive relations of two-fluid model}

Table. \ref{tab:Constitutive} summarizes the corresponding constitutive relations in this work. Note that two frictional stress models are used in different cases and the boundary conditions for the particle phase based on \cite{johnson1987frictional} used in the bubbling fluidized cases are also presented. The calculation method of the granular pressure gradient proposed by \cite{liu2024ps} is used.
Table. \ref{tab:drag} summarizes the three drag models used in this work.

\begin{longtable}[H]{lr}
    \caption{Constitutive relations. \label{tab:Constitutive}}
    \\\hline
    Stress tensor for gas and solid phases \\
    ${\bf \tau_g} = \mu_g[\nabla{\bf u}_g+\nabla{\bf u}_g^T]-\frac23\mu_g(\nabla\cdot{\bf u}_g)\bm I$\\
    ${\bf \tau_s} = \mu_s[\nabla{\bf u}_s+\nabla{\bf u}_s^T]+(\lambda_s-\frac23\mu_s)(\nabla\cdot{\bf u}_s)\bm I$\\
    Solid pressure \citep{lun1984kinetic}\\
    $p_s = \rho_s\alpha_s\left[1+2(1+e)\alpha_s g_0\right]\theta_s$\\
    Frictional pressure \\
    \quad Johnson-Jackson Model \citep{johnson1990frictional} \\
    $p_s^{fr} = \frac{F_r(\alpha_s-\alpha_{s,min})^{eta}}{(\alpha_{s,max}-\alpha_s)^p},\quad F_r=0.05, \quad eta = 2, \quad p=5, \quad \alpha_{s,min}=0.5, \quad \alpha_{s,max}=0.62$\\
    \quad Schaeffer Model \citep{schaeffer1987InstabilityEvolutionEquations}\\
    $p_s^{fr} = 10^{25} (\alpha_s - \alpha_{s,min})^{10}\quad \alpha_{s,min}=0.61$ \\
    Frictional shear stress \citep{schaeffer1987InstabilityEvolutionEquations}\\
    $\mu_s^{fr} = p_s\frac{\sqrt{2}\sin\phi}{2\sqrt{{\bf S_s:S_s}}}
    \quad
    \text{with}
    \quad
    {\bf S_s} = \nabla{\bf u}_s+\nabla{\bf u}_s^T
    \text{,}
    \quad
    \phi = 28.5^{\circ}$ \\
    Granular pressure gradient \citep{liu2024ps}\\
    $\nabla p_s = \frac{\partial p_s}{\partial \alpha_s} \nabla \alpha_s + \frac{\partial p_s}{\partial \theta_s} \nabla \theta_s$ \\
    $\frac{\partial p_s}{\partial \alpha_s} = \rho_s\left[1+\alpha_s(1+e)\left(4g_0+2\frac{\partial g_0}{\partial \alpha_s} \alpha_s\right)\right]\theta_s + Fr\frac{ eta(\alpha_s - \alpha_{s,min})^{eta-1}(\alpha_{s,max}-\alpha_s)+p(\alpha_s-\alpha_{s,min})^{eta} }{(\alpha_{s,max}-\alpha_s)^{p+1}}$ \\
    $\frac{\partial g_0}{\partial \alpha_s}= \frac{1}{ 3 \alpha_{s,max}\left[ \left(\frac{\alpha_s}{\alpha_{s,max}}\right)^{\frac{1}{3}} - \left(\frac{\alpha_s}{\alpha_{s,max}}\right)^{\frac{2}{3}} \right]^2 }$ \\
    $\frac{\partial p_s}{\partial \theta_s} = \alpha_s \rho_s [ 1+2(1+e)\alpha_sg_0(\alpha_s) ]$ \\
    Solid viscosity \citep{syamlal1993mfix}\\
    $\mu_s=\frac{5}{48}\frac{\rho_s d_p \sqrt{\pi\theta_s}}{\alpha_s(1+e)g_0}
        \left[1+\frac45g_0\alpha_s(1+e)\right]^2
        +\frac45\alpha_s^2\rho_s d_p g_0(1+e)\sqrt{\frac{\theta_s}{\pi}}$ &\\
    Radial distribution function \citep{savage1988streaming,sinclair1989gas}\\
    $g_0 = [1-(\frac{\alpha_s}{\alpha_{s,max}})^{\frac{1}{3}}]^{-1}$ \\
    Diffusion coefficient \citep{syamlal1993mfix}\\
    $\kappa_s =
\frac{150\rho_sd_p\sqrt{\theta_s\pi}}{384(1+e)g_0}
\left[1+\frac65\alpha_s g_0(1+e)\right]^2
+2\rho_s\alpha_s^2 d_p(1+e)g_0\sqrt{\frac{\theta_s}{\pi}}$ \\
    Collisional energy dissipation \citep{lun1984kinetic}\\
    $\gamma_s = 12(1-e^2)\alpha_s^2\rho_s g_0\frac{1}{d_p}\sqrt{\frac{\theta_s}{\pi}}$ \\
    Particle volume viscosity \citep{lun1984kinetic}\\
    $\lambda_s = \frac43\alpha_s\rho_s d_p g_0(1+e)\sqrt{\frac{\theta_s}{\pi}}$ \\
    Granular energy equation \citep{gidaspow1994multiphase}\\
    $\frac23\left[
\frac{\partial(\alpha_s\rho_s\theta_s)}{\partial t}
+
\nabla\cdot(\alpha_s\rho_s{\bf u}_s\theta_s)
\right]
-
\nabla\cdot\left(\kappa_s\nabla\theta_s\right)
=
-{p_s{\bf I}}:\nabla{\bf u}_s+{\alpha_s\bf \tau_s}:\nabla{\bf u}_s
-\gamma_s\theta_s
-3\beta\theta_s,$ \\
Johnson-Jackson boundary condition \citep{johnson1987frictional}\\
$\tau_{s,w} = - \frac{\pi}{6}\frac{\alpha_s}{\alpha_{s,max}}\phi \rho_s g_0 \sqrt{3\theta_s}\bm u_{s,w}$ \\
$q_{\theta_s,s} = \frac{\pi}{6} \frac{\alpha_s}{\alpha_{s,max}} \varphi \rho_s g_0 \sqrt{3 \theta_s} |\bm u_{s,w}|^2 - \frac{\pi}{4}\frac{\alpha_s}{\alpha_{s,max}} (1-e^2_{p,w}) \rho_s g_0 \sqrt{3 \theta_s^3}$ \\
    \hline
\end{longtable}

\begin{longtable}[H]{lr}
    \caption{Drag models. \label{tab:drag}}
    \\\hline
    Gidaspow Model \citep{gidaspow1994multiphase} \\
    ${\bf F}_{drag} = \beta({\bf u_g-u_s})
    \quad
    \text{with}
    \quad
    \beta = \begin{cases}
        \frac34 C_D\frac{\rho_g\alpha_g\alpha_s|\bf u_g-u_s|}{d_p}\alpha_g^{-2.65} & \alpha_s\le 0.2\\
        150\frac{\alpha_s^2\mu_g}{\alpha_gd_p^2}+1.75\frac{\rho_g\alpha_s|\bf u_g-u_s|}{d_p} & \alpha_s>0.2
    \end{cases}$ \\
    $C_D = \begin{cases}
        \frac{24}{Re}(1+0.15Re^{0.687}) & Re<1000\\
        0.44 & Re\ge1000
    \end{cases}$\\
    Syamlal-O'Brien Model \citep{syamlal1998mfix} \\
    ${\bf F}_{drag} = \beta({\bf u_g- u_s})
    \quad
    \text{with}
    \quad
    \beta = \frac34 C_D \frac{\rho_g \alpha_g \alpha_s}{V_r^2 d_p}|\bf u_g - u_s|$ \\
    $C_D = \left( 0.63+4.8\sqrt{\frac{V_r}{Re}} \right)^2
    \quad
    \text{with}
    \quad
    V_r = 0.5[ a-0.06Re+\sqrt{(0.06Re)^2+0.12Re(2b-a)+a^2} ]$ \\
    $a=\alpha_g^{4.14},\quad b=\begin{cases}
    0.8\alpha_g^{1.28} & \alpha_g \le 0.85 \\
    \alpha_g^{2.65} & \alpha_g > 0.85
    \end{cases}$\\
    $Re = \frac{\alpha_g\rho_gd_p|\bf u_g-u_s|}{\mu_g}$\\
    EMMS-based drag model \citep{li1994particle,lu2009searching} \\
    ${\bf F}_{drag} = \beta_{EMMS}({\bf u_g-u_s})
    \quad
    \text{with}
    \quad
    \beta_{EMMS} = \begin{cases}
        \frac34 C_D\frac{\rho_g\alpha_g\alpha_s|\bf u_g-u_s|}{d_p}\alpha_g^{-2.65}H_d & \alpha_s\le 0.35\\
        150\frac{\alpha_s^2\mu_g}{\alpha_gd_p^2}+1.75\frac{\rho_g\alpha_s|\bf u_g-u_s|}{d_p} & \alpha_s>0.35
    \end{cases}$ \\
    $H_d = \frac{\beta_{EMMS}}{\beta_{WenYu}} $ \\
    $C_D = \begin{cases}
        \frac{24}{Re}(1+0.15Re^{0.687}) & Re<1000\\
        0.44 & Re\ge1000
    \end{cases}$\\
    \hline
\end{longtable}

\section*{Appendix B: Numerical algorithm of Rhie-Chow momentum interpolation for Euler-Euler gas-solid system}
For two neighboring cells $P$ and $F$ with a face $f$, the $k-$phase velocity for cell $P$ using a staggered grid formulation can be expressed as
\begin{equation}
\begin{split}
&\left(\bm u_{k}\right)_P = \\
&\lambda\left(\frac{\mathbb H_k}{\mathbb D_k}\right)_P + \frac{\lambda}{\Delta t} \left(\frac{\alpha_k \rho_k}{\mathbb D_k}\right)_P \left(\bm u_{k}^i\right)_P + (1 -\lambda)\left(\bm u_{k}^*\right)_P  -\left( \frac{\alpha_{k}}{\mathbb D_k}\right)_P\left(\nabla p_g\right)_P
- \zeta \left(\frac{1}{\mathbb D_k}\right)_P \left(\nabla p_s\right)_P
+ \left(\frac{\alpha_k \rho_k}{\mathbb D_k}\right)_P \bm g
+ \left(\frac{\beta}{\mathbb D_k}\right)_P \left(\bm u^*_{1-k}\right)_P,
\end{split}
\end{equation}
and for cell $F$, we have
\begin{equation}
\begin{split}
&\left(\bm u_{k}\right)_F = \\
&\lambda\left(\frac{\mathbb H_k}{\mathbb D_k}\right)_F + \frac{\lambda}{\Delta t} \left(\frac{\alpha_k \rho_k}{\mathbb D_k}\right)_F \left(\bm u_{k}^i\right)_F + (1 -\lambda)\left(\bm u_{k}^*\right)_F  -\left( \frac{\alpha_{k}}{\mathbb D_k}\right)_F\left(\nabla p_g\right)_F
- \zeta \left(\frac{1}{\mathbb D_k}\right)_F \left(\nabla p_s\right)_F
+ \left(\frac{\alpha_k \rho_k}{\mathbb D_k}\right)_F \bm g
+ \left(\frac{\beta}{\mathbb D_k}\right)_F \left(\bm u^*_{1-k}\right)_F,
\end{split}
\end{equation}
where the subscripts $P$ and $F$ denote the values stored at cell center of $P$ and $F$, respectively.

Therefore, the form of the equation using the Rhie-Chow interpolation is
\begin{equation}
\begin{split}
&\lceil \bm u_{k} \rceil _f = \\
&\lambda\left(\frac{\mathbb H_k}{\mathbb D_k}\right)_f + \frac{\lambda}{\Delta t} \left(\frac{\alpha_k \rho_k}{\mathbb D_k}\right)_f \lceil \bm u_{k}^i \rceil _f + (1 -\lambda) \lceil \bm u_{k}^* \rceil _f
-\left( \frac{\alpha_{k}}{\mathbb D_k}\right)_f \lceil \nabla p_g \rceil _f
- \zeta \left(\frac{1}{\mathbb D_k}\right)_f \lceil\nabla p_s \rceil _f
+ \left(\frac{\alpha_k\rho_k}{\mathbb D_k}\right)_f \bm g
+ \left(\frac{\beta}{\mathbb D_k}\right)_f \lceil \bm u^*_{1-k} \rceil _f,
\label{RCuf}
\end{split}
\end{equation}

The average of the first term on the right hand side of Eq. (\ref{RCuf}) can be obtained according to Eq. (\ref{linear}):
\begin{equation}
\begin{split}
 \lambda \left(\frac{\mathbb H_k}{\mathbb D_k}\right)_f & =
\lambda \omega \left(\frac{\mathbb H_k}{\mathbb D_k}\right)_P+\lambda(1-\omega) \left( \frac{\mathbb H_k}{\mathbb D_k}\right)_F \\
&=(\bm u_k)_f - \frac{\lambda}{\Delta t} \left(\frac{\alpha_k \rho_k}{\mathbb D_k}\right)_f \left(\bm u_k^i\right)_f - (1 -\lambda) \left(\bm u_k^*\right)_f
+\left( \frac{\alpha_{k}}{\mathbb D_k}\right)_f\left(\nabla p_g\right)_f
+ \zeta \left(\frac{1}{\mathbb D_k}\right)_f \left(\nabla p_s\right)_f - \left(\frac{\alpha_k \rho_k}{\mathbb D_k}\right)_f \bm g
- \left(\frac{\beta}{\mathbb D_k}\right)_f \left(\bm u^*_{1-k}\right)_f.
\label{RCH}
\end{split}
\end{equation}
Substituting Eq. (\ref{RCH}) into Eq. (\ref{RCuf}), the Rhie-Chow interpolated cell face velocity $\lceil \bm u_{k} \rceil _f$ is obtained as
\begin{equation}
\begin{split}
\lceil \bm u_{k} \rceil _f = (\bm u_k)_f+ \frac{\lambda}{\Delta t}\left(\frac{\alpha_k \rho_k}{\mathbb D_k}\right)_f\left[\lceil \bm u_{k}^i \rceil _f - (\bm u^i_k)_f\right] + (1 -\lambda)\left[\lceil \bm u_{k}^* \rceil _f - (\bm u^*_k)_f\right]\\
-\left( \frac{\alpha_{k}}{\mathbb D_k}\right)_f \left[\lceil \nabla p_g \rceil _f- \left( \nabla p_g \right)_f \right]- \zeta \left(\frac{1}{\mathbb D_k}\right)_f\left[ \lceil \nabla p_s \rceil_f - \left( \nabla p_s \right)_f \right]
+\left(\frac{\beta}{\mathbb D_k}\right)_f\left[ \lceil \bm u_{1-k}^* \rceil _f - (\bm u_{1-k}^*)_f\right].
\label{RCukf}
\end{split}
\end{equation}

There are several crucial aspects to consider regarding the Rhie-Chow correction in OpenFOAM$^{\circledR}$ in this study:

(i)The implementation of transient Rhie-Chow correction term $\frac{\lambda}{\Delta t}\left(\frac{\alpha_k \rho_k}{\mathbb D_k}\right)_f\left[\lceil \bm u_{k}^i \rceil _f - (\bm u^i_k)_f\right]$ (also called Choi correction) depends on the employed time-discretization scheme. For example, if solving the transient term with a second-order backward scheme:
\begin{equation}
\frac{\partial }{\partial t}\left(\alpha_k\rho_k\bm u_k\right) = \frac{3\alpha_k\rho_k\bm u_k -4\alpha_k^i\rho_k\bm u_k^i+\alpha_k^{i-1}\rho_k\bm u_k^{i-1}}{2\Delta t}.
\end{equation}
Then the transient Rhie-Chow correction term $\frac{\lambda}{\Delta t}\left(\frac{\alpha_k \rho_k}{\mathbb D_k}\right)_f \left[\lceil \bm u^i_{k} \rceil _f - (\bm u^i_k)_f\right]$ in Eq. (\ref{RCukf}) originated by \citet{choi1999note} should be substituted by $\frac{\lambda}{2\Delta t}(\frac{1}{\mathbb D_k})_f\left[ 4(\alpha_k^i \rho_k)_f\left( \lceil \bm u^i_{k} \rceil _f - (\bm u^i_k)_f \right)-(\alpha_k^{i-1} \rho_k)_f\left( \lceil \bm u^{i-1}_{k} \rceil _f - (\bm u^{i-1}_k)_f \right) \right]$ \citep{shen2001improved,cubero2007compact}.
The transient correction term may appear to be insignificant in most practical computations, yet acquiring a solution that is dependent on the time step is not desired.
However in the OpenFOAM$^{\circledR}$ code, an additional limiter $\gamma \in [0,1]$ has been added to this term. It is an undocumented additional empirical coefficient and is not present in the original formulation of \citet{choi1999note}, which is defined as
\begin{equation}
\gamma = 1 - min\left[ \frac{|\varphi_{k}^i - (\bm u_k^i)_f \cdot \bm S|}{|\varphi_{k}^i|+ \epsilon},1 \right],
\end{equation}
where $\varphi_{k}^i=\lceil \bm u_{k}^i \rceil _f \cdot \bm S$ is the conservative face flux at the previous time step, $\epsilon$ is a small number to prevent the occurrence of division by zero. By default, $\gamma$ is set to 1, at this time, the transient correction term is the full Choi correction. However, this term may induce instability in certain circumstances, prompting the empirical $\gamma$ limiter to be employed as a means of constraining the aforementioned Choi correction. \citet{vuorinen2014implementation} investigated the impact of this transient Rhie-Chow correction term and concluded that it may contribute to numerical diffusivity in certain scenarios. Moreover, researches conducted by \cite{bartholomew2018unified} and \cite{xiao2017fully} proved that using the momentum interpolation method without transient correction can cause a dispersion error in pressure signals.\\

(ii) The term $(1 -\lambda)\left[\lceil \bm u_{k}^* \rceil _f - (\bm u^*_k)_f\right]$ in Eq. (\ref{RCukf}) is ordinarily accepted as the Majumdar correction \citep{majumdar1988role}, which can efficiently eradicate the dependency on under-relaxation factor with no evident impact on convergence rate \citep{martinez2017influence}.

(iii)The term $-\left( \frac{\alpha_{k}}{\mathbb D_k}\right)_f \left[\lceil \nabla p_g \rceil _f- \left( \nabla p_g \right)_f \right]$ is the gas pressure Rhie-Chow correction term. When solving for gas pressure Poisson equation at the face, the value of gas pressure surface-normal gradient $\nabla^{\perp} p$ stored at the cell face is used to substitute the interpolated gas pressure gradient $\lceil \nabla p_g \rceil_f$ in Eq. (\ref{RCukf}), which defines as
\begin{equation}
\lceil \nabla p_g \rceil_f = \nabla^{\perp}p_g = \frac{p_{g,P}-p_{g,N}}{|\bm d_P -\bm d_N|}
\end{equation}
where subscript $P$ and $N$ represents the neighbouring cells of the current face, and $\bm d$ is the cell-centered coordinate vector.
\begin{figure}[!htb]
    \centering
    \includegraphics[scale=1]{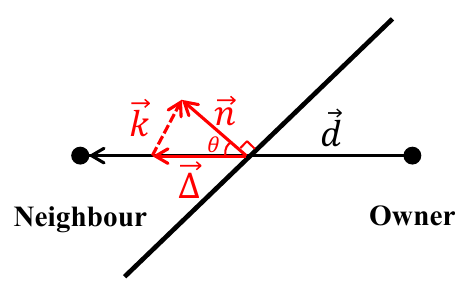}
    \caption{Decomposition schematic of the face normal vector.}
    \label{nonorthogonal}
\end{figure}
There is a critical problem when solving the gas pressure equation. When the geometric meshes are not completely orthogonal, as shown in Fig. \ref{nonorthogonal}, where $\vec d$ is the vector pointing from the center of Owner cell to the center of Neighbour cell. The surface normal gradient of the pressure $\nabla^{\perp}p_g$ in the direction of $\vec n$ can be decomposed into the orthogonal component $\vec \Delta$ parallel to $\vec d$ and the non-orthogonal component $\vec k$
\begin{equation}
\vec n = \vec \Delta + \vec k.
\end{equation}
The value of the orthogonal component $\vec \Delta$ depends on the decomposition method. In this study, the over-relaxed approach is used \citep{greenshieldsweller2022}, since it is able to decompose the non-orthogonal part to a minimum and works best compared to other approaches \citep{jasak1996error}
\begin{equation}
\vec \Delta = \frac{|\vec n|}{cos \theta} \frac{\vec d}{|\vec d|},
\end{equation}
where $\theta$ is the angel between $\vec \Delta$ and $\vec n$. The non-orthogonal correction is required to provide second-order accuracy and guarantee the low-pass filter characteristic on pressure field. Then the gradient of gas pressure at the face becomes:
\begin{equation}
\nabla^{\perp} p_g = \vec \Delta \frac{p_{g,P}-p_{g,N}}{|\bm d_P - \bm d_N|}+\vec k (\nabla p_g)_f
\label{equ:nonorthogonal}
\end{equation}
where the surface interpolation of $(\nabla p_g)_f$ is calculated as same as Eq. (\ref{linear}). The implicit computation of the orthogonal term $\vec \Delta \frac{p_{g,P}-p_{g,N}}{|\bm d_P - \bm d_N|}$ in Eq. (\ref{equ:nonorthogonal}) includes only the first neighbors of the cell and would be added in the coefficient matrix,  resulting in a diagonally equal matrix. In contrast, implicitly dealing with the non-orthogonal correction term $\vec k (\nabla p_g)_f$ in Eq. (\ref{equ:nonorthogonal}) would introduce the "second neighbors" of the control volume into the computational molecule with negative coefficients, violating diagonal equality, and possibly causing unboundedness, especially when the mesh non-orthogonality is high. If ensuring boundedness is of priority over accuracy, it may be necessary to limit or completely eliminate the non-orthogonal correction term \citep{rusche2003computational}. Therefore, the non-orthogonal correction is explicitly treated in this work and is added into the corresponding source vector \citep{greenshieldsweller2022}. On the other hand, the non-orthogonal correction can provide considerable enhancement in precision and robustness \citep{bartholomew2018unified}.
Therefore, the gas pressure gradient of the current cell is only discretized and calculated using the consecutive cells instead of alternating cells. As a result, the final matrix is a compact arrangement, avoiding the checkerboard pressure field and enhancing the convergence of numerical solution.

(iv)The term $- \zeta \left(\frac{1}{\mathbb D_k}\right)_f\left[ \lceil \nabla p_s \rceil_f - \left( \nabla p_s \right)_f \right]$ is the granular pressure Rhie-Chow correction term. To avoid saddle point problem for granular pressure gradient, $\lceil \nabla p_s \rceil _f$ is substituted by $\nabla^{\perp}p_s$ \citep{liu2024ps}:
\begin{equation}
\left(\nabla p_s\right)_f = \nabla^{\perp}p_s = \frac{\partial p_s}{\partial \alpha_s} \nabla^{\perp}\alpha_s + \frac{\partial p_s}{\partial \theta_s} \nabla^{\perp}\theta_s = \frac{\partial p_s}{\partial \alpha_s}\left(\vec \Delta \frac{\alpha_{s,P}-\alpha_{s,N}}{|\bm d_P -\bm d_N|} + \vec k (\nabla \alpha_s)_f\right) + \frac{\partial p_s}{\partial \theta_s} \left( \vec \Delta \frac{\theta_{s,P}-\theta_{s,N}}{|\bm d_P -\bm d_N|} + \vec k (\nabla \theta_s)_f \right)
\end{equation}
The non-orthogonal modification enables the momentum interpolation method to be applied to arbitrary meshes.
\clearpage
\section*{Reference}

\end{CJK*}
\end{sloppypar}
\end{document}